\documentclass[11pt,a4paper]{article}
\usepackage[utf8]{inputenc}
\usepackage[T1]{fontenc}
\usepackage{amsmath,amssymb,amsthm,mathtools, mathrsfs}
\usepackage{graphicx,xcolor,booktabs,placeins,needspace}
\usepackage{MorrisIn,lettrine,Romantik}

\usepackage[margin=1in]{geometry}
\usepackage{authblk}
\usepackage[colorlinks=true,linkcolor=blue!50!black,citecolor=blue!50!black,urlcolor=blue!50!black]{hyperref}
\usepackage{tikz}

\usetikzlibrary{arrows.meta,calc}
\definecolor{holo}{HTML}{1768AC}
\definecolor{anti}{HTML}{C16A28}
\definecolor{surface}{HTML}{DCE6EF}
\definecolor{torusdark}{HTML}{6D879E}
\definecolor{toruslight}{HTML}{F1F6FA}
\newcommand{\Young}[2]{%
  \foreach \row [count=\i from 0] in {#1}{%
    \foreach \j in {1,...,\row}{%
      \draw[draw=#2,line width=.65pt,fill=#2!9]
        ({(\j-1)*.25},{-\i*.25}) rectangle ++(.25,-.25);
    }%
  }%
}

\theoremstyle{plain}
\newtheorem{lemma}{Lemma}[section]
\newtheorem{proposition}[lemma]{Proposition}
\theoremstyle{definition}
\newtheorem{definition}[lemma]{Definition}

\newcommand{\dd}{\mathrm{d}}
\newcommand{\cF}{\mathcal{F}}
\newcommand{\cN}{\mathcal{N}}

\newcommand{\cS}{\mathcal{S}}
\newcommand{\cC}{\mathcal{C}}
\newcommand{\bz}{\bar z}
\newcommand{\bq}{\bar q}
\newcommand{\Th}{\Theta}

\renewcommand{\Re}{\operatorname{Re}}
\renewcommand{\Im}{\operatorname{Im}}

\title{\vspace{-1cm}Complex self-energy matrix at one loop for massive states in Type II superstrings}

\author[1]{Diptarka Das\thanks{\href{mailto:diptarka.das@gmail.com}{diptarka.das@gmail.com}}}
\author[1]{Santanu Mandal\thanks{\href{mailto:santanum20@iitk.ac.in}{santanum20@iitk.ac.in}}}
\author[2]{Anurag Sarkar\thanks{\href{mailto:anurag.sarkar@wits.ac.za}{anurag.sarkar@wits.ac.za}}}
\author[1]{Aleena Grace Tojan\thanks{\href{mailto:aleenagt26@iitk.ac.in}{aleenagt26@iitk.ac.in}}}
\author[3]{Mritunjay Verma\thanks{\href{mailto:mritunjay@iiti.ac.in}{mritunjay@iiti.ac.in}}}
\affil[1]{Department of Physics, Indian Institute of Technology Kanpur, Kanpur 208016, India}
\affil[2]{National Institute for Theoretical and Computational Sciences, School of Physics, and Mandelstam Institute for Theoretical Physics, University of the Witwatersrand, Johannesburg, Wits 2050, South Africa.}
\affil[3]{Department of Physics, Indian Institute of Technology Indore, Indore 453552, India }

\date{}

\begin{document}

\begin{titlepage}

\maketitle

\begin{abstract}
We compute the one-loop complex self-energy matrix within a family of
massive DDF states in Type II superstring theory. The states are obtained
by acting on a massless seed with bosonic DDF operators of fixed transverse
polarization. The key ingredient is a finite expansion of the spin-summed
chiral blocks in a generalized theta basis. It allows us to carry out the
worldsheet insertion integral exactly, leaving a modular integral whose
long-torus contribution gives the absorptive part after analytic continuation.
We apply this construction to bosonic partitions through level $K=10$.
At low levels we recover the complex pole shifts of all three long
supermultiplets at $K=2$ and of six multiplets at $K=3$. With the larger
matrices we ask how the spacings of the eigenvalues of the mass-shift
matrix depend on the shapes of the partitions. Our results indicate that typical shapes yield random matrix distributed mass shifts.
\end{abstract}
\end{titlepage}

\vfill

\vfill \eject

\baselineskip18pt

\tableofcontents



\section{Introduction}
\begin{figure}[htbp]
\centering
\begin{tikzpicture}[
    >=Latex,
    thick,
    font=\small
]

\begin{scope}
    \node at (1.5,3.3) {\textbf{Tree level}};

    \draw[->] (0,0) -- (0,2.9) node[above] {$M^2$};
    \draw[->] (0,0) -- (3,0);

    \draw[blue,very thick] (0.45,1.65) -- (2.55,1.65);
    \node[left] at (0,1.65) {$M_{N,0}^2$};

    \node[align=center] at (1.5,-0.55)
    {One mass, many states};
\end{scope}

\draw[->,thick] (3.4,1.6) -- (4.2,1.6);
\node[above] at (3.8,1.65) {$g_s\neq 0$};

\begin{scope}[xshift=4.7cm]
    \node at (2.5,3.3) {\textbf{Including loops}};

    \draw[->] (0,0) -- (0,2.9) node[above] {$M^2$};
    \draw[->] (0,0) -- (3,0);

    \foreach \y in {1.10,1.25,1.42,1.58,
                     1.73,1.90,2.08,2.25,2.40}
    {
        \draw[teal,thick] (0.45,\y) -- (2.35,\y);
    }

    \draw[<->,thin] (2.6,1.10) -- (2.6,2.40);
    \node[right] at (2.65,1.75) {$\Delta M^2$};

    \node[align=center] at (1.5,-0.55)
    {Split mass eigenvalues};
\end{scope}

\end{tikzpicture}
\caption{Schematic illustration of the lifting of
tree-level mass degeneracy by string interactions.
}
\label{fig:mass_splitting}
\end{figure}
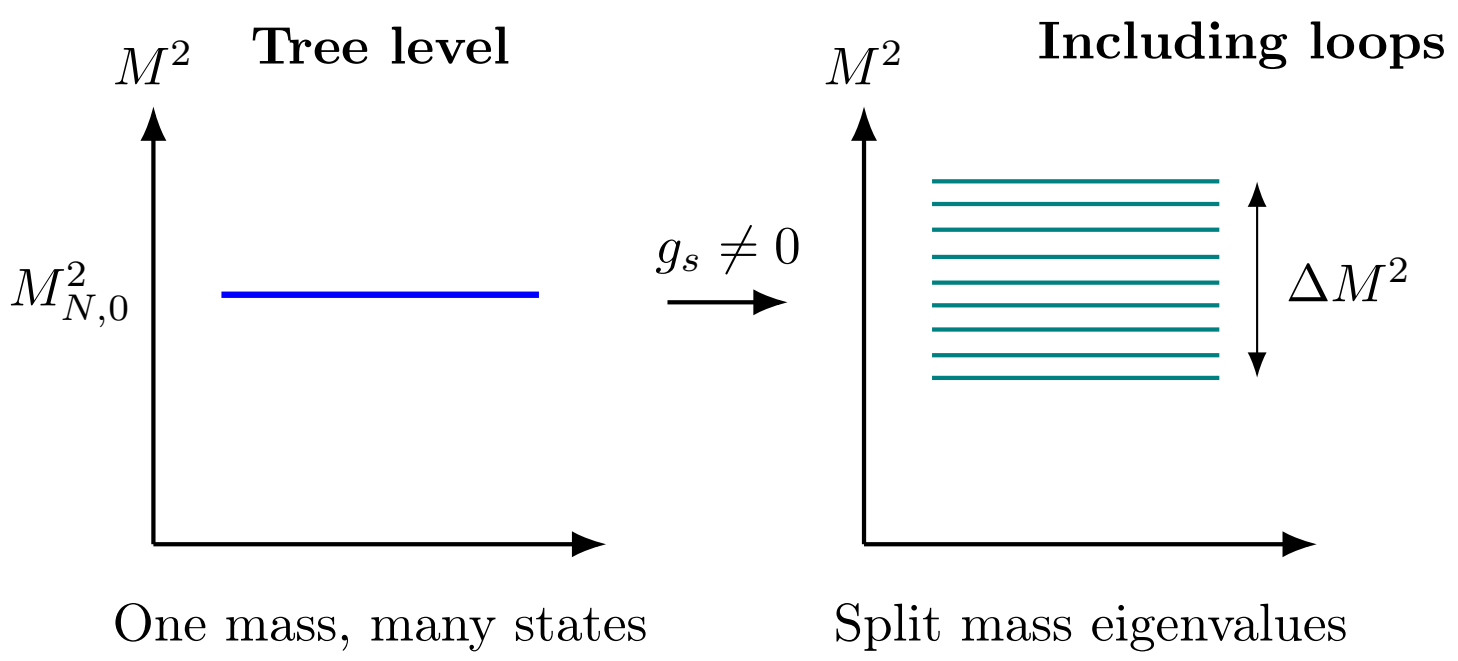

\lettrine[lines=3]{Q}{uantum\,gravity} becomes calculable order by order
in perturbative closed string theory. Alongside the graviton, the spectrum
contains infinitely many massive spinning states. At tree level, the mass
squared is fixed by the number of oscillator excitations counted with their
mode numbers. There are many ways to distribute these excitations, and the
number of states grows exponentially in the square root of the level.
At nonzero string coupling the degenerate states can mix. Finding the
corrected masses is then a matrix problem. To leading
order, the inverse propagator projected onto the physical states of a
given tree-level mass $M_0$ takes the form
\begin{equation}
 G^{-1}_{ab}(p)=(p^2+M_0^2)\delta_{ab}+\Sigma_{ab}(p^2)
                 +\mathcal O(g_s^4),
\end{equation}
where $\Sigma$ includes the factor of $g_s^2$. To find the corrected
masses we have to diagonalize $\Sigma(-M_0^2)$ within the degenerate
space. The eigenvalues are the order-$g_s^2$ pole shifts $\Delta M_r^2$.
Their real parts can spread the original mass level into a band, as in
Fig.~\ref{fig:mass_splitting}, with degeneracies within each irreducible
symmetry multiplet protected. There is another consequence of the loop
correction. A massive string can decay into lighter strings, and when
such channels are open the pole is complex,
\begin{equation}
 M_{{\rm pole},r}^2=M_0^2+\Delta M_r^2
 =M_0^2+2M_0\,\delta M_r-iM_0\Gamma_r+\mathcal O(g_s^4).
\end{equation}
The width $\Gamma_r$ is the inverse lifetime at this order.

\begin{figure}[htbp]
\centering
\begin{tikzpicture}[>=Latex,font=\small,line cap=round,line join=round]
  \shade[top color=surface!15,bottom color=torusdark!40!surface]
    (0,0) ellipse (2.75 and 1.50);
  \draw[black,line width=.9pt] (0,0) ellipse (2.75 and 1.50);
  \filldraw[fill=white,draw=black,line width=.85pt]
    (-1.30,-.06)
      .. controls (-1.52,.88) and (1.52,.88) .. (1.30,-.06)
      .. controls (.65,-.62) and (-.65,-.62) .. cycle;
  \draw[black,line width=.85pt]
    (-1.50,.28) .. controls (-1.49,.14) and (-1.42,.04) .. (-1.30,-.06);
  \draw[black,line width=.85pt]
    (1.30,-.06) .. controls (1.42,.04) and (1.49,.14) .. (1.50,.28);
  \node[font=\normalsize] at (0,1.85) {Torus world-sheet $\Sigma_\tau$};

  \coordinate (insertionA) at (-1.98,-.47);
  \coordinate (insertionB) at (1.98,.52);
  \draw[-{Latex[length=2.5mm,width=1.7mm]},black!80,line width=.85pt,
    preaction={draw=white,line width=2.7pt},shorten >=4pt]
    (-3.97,-.47) -- (insertionA);
  \draw[-{Latex[length=2.5mm,width=1.7mm]},black!80,line width=.85pt,
    preaction={draw=white,line width=2.7pt},shorten <=4pt]
    (insertionB) -- (3.97,.52);
  \fill (insertionA) circle (2.3pt);
  \fill (insertionB) circle (2.3pt);
  \node[below=5pt] at (insertionA) {$0$};
  \node[below=5pt] at (insertionB) {$z$};

  \begin{scope}[xshift=-5.65cm]
    \node[font=\large] at (0,2.24) {$|\alpha,\alpha'\rangle$};
    \node[text=holo] at (0,1.55) {Holomorphic};
    \begin{scope}[shift={(-.625,1.27)}] \Young{5,3,1}{holo} \end{scope}
    \node at (0,.23) {$\alpha=(5,3,1)$};
    \node[text=anti] at (0,-.32) {Antiholomorphic};
    \begin{scope}[shift={(-.5,-.60)}] \Young{4,3,2}{anti} \end{scope}
    \node at (0,-1.64) {$\alpha'=(4,3,2)$};
  \end{scope}

  \begin{scope}[xshift=5.65cm]
    \node[font=\large] at (0,2.24) {$\langle\gamma,\gamma'|$};
    \node[text=holo] at (0,1.55) {Holomorphic};
    \begin{scope}[shift={(-.75,1.27)}] \Young{6,2,1}{holo} \end{scope}
    \node at (0,.23) {$\gamma=(6,2,1)$};
    \node[text=anti] at (0,-.32) {Antiholomorphic};
    \begin{scope}[shift={(-.5,-.60)}] \Young{4,4,1}{anti} \end{scope}
    \node at (0,-1.64) {$\gamma'=(4,4,1)$};
  \end{scope}

  \node[font=\normalsize] at (0,-2.35)
    {$\displaystyle
      A_{\gamma\gamma',\alpha\alpha'}(z,\bar z\mid\tau)
      =\bigl\langle V^{\dagger}_{\gamma,\gamma'}(z,\bar z)\,
                       V_{\alpha,\alpha'}(0,0)\bigr\rangle_{\Sigma_\tau}$};
  \node at (0,-3.02)
    {$|\alpha|=|\alpha'|=|\gamma|=|\gamma'|=K=9,
       \qquad \alpha'_{\!\rm string}M_0^2=4K=36$};
\end{tikzpicture}
\caption{Torus two-point function between pairs of DDF partitions. The example
has $K=9$ in each chiral sector. Translation invariance relates this placement
of the two vertices to the convention used in Section~\ref{sec:one-loop}.}
\label{fig:torus-ddf}
\end{figure}
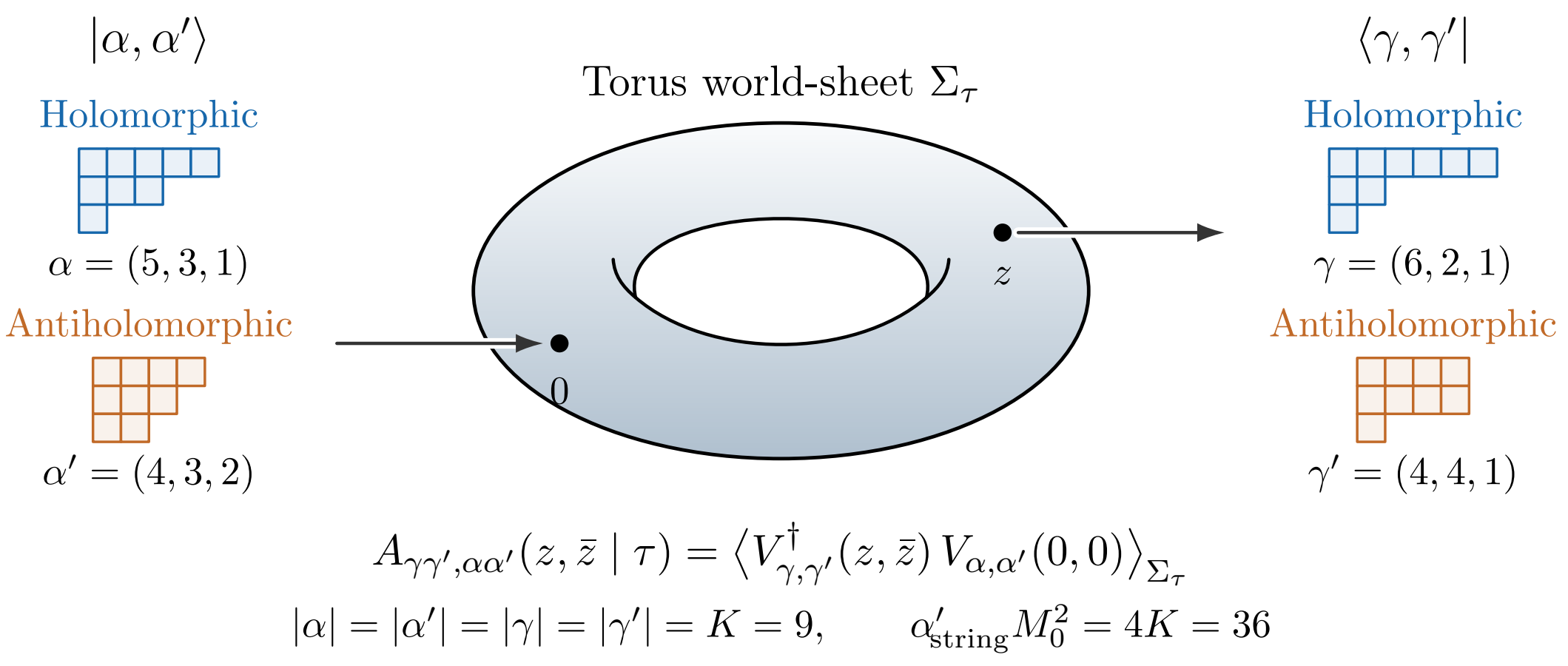

The self-energy diagram is computed on the worldsheet torus. Translation
invariance lets us fix one vertex at the origin, leaving the position $z$
of the other to be integrated over the surface. We then have to sum over
the tori themselves by integrating the modular parameter $\tau$.
Figure~\ref{fig:torus-ddf} shows the two-point function between two pairs
of DDF partitions. The two integrations play rather different roles.
We can do the integral over $z$ exactly, while the imaginary part arises
from continuing the integral over the long-torus region.

In slightly more detail, an excited state $|\alpha,\alpha'\rangle$ is
obtained from a massless seed by acting with one bosonic DDF operator
$A_{-n}(\lambda)$ for each part $n$ of $\alpha$ in the left-moving sector,
and with the corresponding operators for $\alpha'$ on the right. The two
partitions are chosen independently, subject to $|\alpha|=|\alpha'|=K$, since otherwise the correlator vanishes.
There are therefore $p(K)^2$ states of the same tree-level mass squared
$M_0^2=2K$ in units $\alpha'=2$, where $p(K)$ is the number of partitions
of $K$. The self-energy matrix is related to the reduced, integrated
amplitude by
\begin{align}
 \langle\gamma,\gamma'|\Sigma|\alpha,\alpha'\rangle
 &= -\cN A_{\gamma\gamma',\alpha\alpha'},\qquad \cN>0.
 \label{eq:DeltaM2intro}
\end{align}
Unitarity relates the absorptive part to the tree-level decay amplitudes,
\begin{equation}
 \cN\Im A_{ii}
 =M_0\Gamma_i^{\mathrm{tree}}
 =\frac12\sum_f\int\dd\Phi_f\,
 \bigl|\mathcal M^{\mathrm{tree}}_{i\to f}\bigr|^2,
\end{equation}
where the sum includes the phase-space symmetry factors. Matching this relation
with consistently normalized tree amplitudes fixes $\cN$. We use
$\cN=\pi^2 g_s^2$ at $\alpha'_{\text{string}}=2$ throughout. A complex pole shift determines
the mass shift and width through
\begin{align}
 \delta M&=\frac{\Re\Delta M^2}{2M_0},\qquad
 \Gamma=-\frac{\Im\Delta M^2}{M_0}.
 \label{eq:dMGammaintro}
\end{align}

Two properties of the matrix will be used throughout. The first is
conservation of the rotation charge $J_{23}=\ell(\alpha)+\ell(\alpha')$,
where $\ell$ counts the parts of a partition. The two lengths can change
separately as long as their sum is conserved. The second is left--right
exchange symmetry, which splits each charge block into an even and an
odd part. The matrix is complex symmetric in our phase convention.
These symmetries make the calculation within the DDF family tractable.
The link to physical pole shifts requires the full symmetry sectors, or
enough matrix elements to reconstruct the shifts from their projectors.

The rest of the paper is organized as follows. In Section~2 we construct the family of massive NS--NS states studied in this paper by acting on a massless seed with bosonic DDF operators of fixed polarization, and give their zero-picture vertex operators. In Section~\ref{sec:one-loop} we compute the one-loop self-energy matrix from the torus two-point function. We expand the spin-summed chiral blocks in a finite theta basis, perform the integral over the vertex position exactly, and evaluate the remaining modular integral. In Section~\ref{sec:results} we present the complex pole shifts at low levels and study the eigenvalue spacings of the mass-shift matrix in sectors of fixed charge and exchange parity through level $K=10$. Appendix~\ref{sec:explicitV} gives explicit vertices at $K=1,2$ and works out the $K=1$ amplitude. Appendix~\ref{proofs} collects the spin-sum and theta identities and gives the modular properties of the integral. Appendix~\ref{app:numdet} describes the numerical methods and error checks. In appendix ~\ref{app:rademacher} we use the Rademacher method to evaluate the same $\tau,\bar{\tau}$ integrals. Appendix~\ref{subsec:ddf_spin_degeneracy} explains how the DDF matrix elements determine the low-level supermultiplet pole shifts, and Appendix~\ref{app:spurious-pco} shows that the amplitude is independent of the positions of the picture-changing operators.

\subsection*{Note : } We choose simplest fermionic content in our states, and only bosonic parititions for ease of computation and illustration of general features. The technology presented here can straightforwardly be generalized to arbitrary states in the DDF basis.

\subsection{Relation to prior work}
\label{sec:prior-work}

Del Giudice, Di Vecchia and Fubini constructed the physical oscillator
basis for the bosonic string~\cite{DDF1972}. The NS and Ramond DDF vertex
operators and their coherent-state versions provide a covariant description
of massive superstring states~\cite{AldiFirrotta2019,BasileMarkou2024}.
The little-group characters and long-multiplet organization of the Type II
spectrum are known~\cite{BianchiMoralesSamtleben2003,ForcellaHananyTroost2010}.
We use the partition basis to evaluate the two-point functions and use the
multiplet decomposition at low levels to extract pole shifts.

Amano and Tsuchiya computed on-shell two-point functions for low-lying
Type II states and found evidence for splitting at the second massive
level~\cite{AmanoTsuchiya1989}. Analytic continuation of genus-one amplitudes
and the relation of their discontinuities to massive decay rates were studied
by D'Hoker and Phong~\cite{DHokerPhong1993,DHokerPhong1994}. Subsequent
calculations obtained widths for maximum-spin states and for families of
long-lived states~\cite{IengoKalkkinen2000,IengoRusso2002,ChialvaIengoRusso2003,
ChialvaIengoRusso2005}. Tree-level calculations have also related inclusive
emission and absorption to the spectrum of available string states
\cite{AmatiRusso1999,FirrottaKiritsisNiarchos2025}. The definition of the renormalized poles for general
massive states and the treatment of unstable states were developed in
Refs.~\cite{PiusRudraSen2014a,PiusRudraSen2014b,Sen:2016gqt}.  Bianchi, Firrotta and Grimaldi recently integrated the insertion position
analytically for states on the leading Regge trajectory and computed their
one-loop mass shifts and widths through level ten~\cite{BianchiFirrottaGrimaldi2026}.
They also raised the possibility of random-matrix statistics for lower-spin
mixing.

 In this work we develop a technique to sum the spins exactly and expand the resulting blocks in a generalized theta basis. Thereafter we apply the $i\varepsilon$ prescription to the long-torus integral
~\cite{Witten2013,Sen2016Contour,EberhardtMizera2022}. We also obtain equivalent answers using the {\em faster converging} Rademacher method \cite{BacciantiEtAl2025} that we adapted to the setting of massive states possessing arbitrary bosonic partitions. Tree-level studies of excited strings have used the angular dependence of
amplitudes and scattering-matrix observables to test random-matrix predictions
\cite{GrossRosenhaus2021,BianchiEtAl2022,BianchiEtAl2023,DasMandalSarkar2024}.
Here we can ask a related question about the spacings of the levels
themselves as they split at one loop. We study the eigenvalues of the
mass-shift matrix in each DDF block and use the choice of partitions
to see how sensitive the spacing distribution is to the states included
in the calculation.

\subsection{Notation}
We compute everything in units of the string length, which is set to $\ell_s=\sqrt{2}$ and use the metric
$\eta_{\mu\nu}=\mathrm{diag}(-1,+1,\dots,+1)$, so that $p^2=-M^2$.
The holomorphic coordinate fields and NS fermions have the plane OPEs
\begin{align}
  X^\mu(z)\,X^\nu(w) &\sim -\eta^{\mu\nu}\log (z-w) , \qquad  \psi^\mu(z)\,\psi^\nu(w) \sim \frac{\eta^{\mu\nu}}{z-w} ,
\end{align}
For a constant vector $v^\mu$ and an integer $d\ge 0$ we write
\begin{align}
  B_d(v) &:= i\, v\cdot \partial^d X ,\qquad F_d(v) := v\cdot \partial^d \psi  .
  \label{eq:BFdef}
\end{align}
The field $B_d(v)$ has conformal weight $d$ and the field $F_d(v)$ has conformal
weight $d+\tfrac12$ for the fields used below.

The torus $T_\tau$ has lattice $\mathbb Z+\tau\mathbb Z$, with
$\tau=\tau_1+i\tau_2$ and $\tau_2>0$. We write $q=e^{2\pi i\tau}$
when $q$ is the nome \footnote{In a slight abuse of notation we use the same letter in the DDF construction the same letter denotes a
null vector.} The fundamental domain $\cF$ has $|\tau_1|\leq\tfrac12$ and
$|\tau|\geq1$. We parameterize the insertion by
$z=\sigma_1+\tau\sigma_2$ with $0\le\sigma_1,\sigma_2<1$, and set
\begin{equation}
  y := \frac{\Im z}{\tau_2} = \sigma_2 ,
  \qquad
  h := \frac{\pi}{\tau_2} .
  \label{eq:yh}
\end{equation}
The area element is $\dd^2 z := \dd(\Re z)\,\dd(\Im z) = \tau_2\,\dd\sigma_1\,\dd\sigma_2$,
and $\dd^2\tau := \dd\tau_1\,\dd\tau_2$.

The torus scalar propagator~\cite{DHokerPhong1988} is
$\langle X^\mu(z,\bz)X^\nu(0)\rangle=\eta^{\mu\nu}G(z)$, with
\begin{equation}
  G(z) = -\log\left|\frac{\theta_1(z|\tau)}{\theta_1'(0|\tau)}\right|^2
         + \frac{2\pi\,(\Im z)^2}{\tau_2} .
  \label{eq:G}
\end{equation}
It is doubly periodic, $G(z+1)=G(z+\tau)=G(z)$, and behaves as $-\log|z|^2$ near
$z=0$. The fermion propagator in the even spin structure
$s\in\{2,3,4\}$ is the Szeg\H{o} kernel
\begin{equation}
  S_s(z) = \frac{\theta_s(z|\tau)\,\theta_1'(0|\tau)}{\theta_s(0|\tau)\,\theta_1(z|\tau)} ,
  \qquad S_s(z) = \frac1z + O(z) .
  \label{eq:Szego}
\end{equation}
We use the genus-one prime form
\begin{equation}
 E(z):=\frac{\theta_1(z|\tau)}{\theta_1'(0|\tau)}
      =z+O(z^3).
 \label{eq:Edef}
\end{equation}
At fixed $\bz$ we define
\begin{equation}
 G_n=\partial_z^n G,\qquad
 L_n=-\partial_z^n\log\theta_1(z|\tau),\qquad n\geq1.
\end{equation}
Thus $G_1=L_1-2\pi i y$, $G_2=L_2-h$, and $G_n=L_n$ for $n\geq3$.
The regular parts at coincidence are
\begin{equation}
 R_n=U_n-h\delta_{n,2},\qquad
 U_n=-\left.\partial_t^n\log\frac{\theta_1(t|\tau)}
                            {\theta_1'(0|\tau)t}\right|_{t=0}.
 \label{eq:Rn}
\end{equation}
For the fermions we use $S_{s,n}=\partial_z^n S_s$ and
$T_{s,n}=\lim_{z\to0}\partial_z^n[S_s(z)-1/z]$.
The mixed contraction of first bosonic derivatives is proportional to $h$.

\section{Highly excited strings in Type II}
We will consider the family of states in the NS-NS sector of Type II closed superstring. A general description of the physical states in the theory can be given by the successive action of the DDF operators on the physical states at any given level. The bosonic DDF operators are constructed from the massless vector vertex operator. Let  $\lambda$ and $q$ be two null vectors satisfying $\lambda.q=0$ and normalization $\lambda.\lambda^{*}=1$. Then the appropriate massless vertex operator in the picture number $(-1,-1)$ is :

\begin{equation}
\mathcal{V}^{-1,-1}
=
 e^{-\phi-\tilde{\phi}}
\left(\lambda.\psi\right)(\lambda.\tilde{\psi})\ 
e^{-iq\cdot X}.
\
\end{equation}
Here $\tilde{\psi}$ is the right moving fermion and $\psi$ is the left moving fermion.The fields $\phi$ and $\tilde{\phi}$ are the bosonised superghost and the same polarization $\lambda$ is used in the both chiral halves. The integrated vertex operator in the picture number (0,0) can be obtained through the action of $G_{-\frac{1}{2}}$ the mode of worldsheet supercurrent. The action of this operator increases the superconformal weight by a unit of $\frac{1}{2}$.
\begin{equation}
    \mathcal{V}^{0,0}=G_{-\frac{1}{2}}\tilde{G}_{{-\frac{1}{2}}}\left(\lambda.\psi\right)(\lambda.\tilde{\psi})\ 
e^{-iq\cdot X}.
\end{equation}
The integrated vertex operator for the holomorphic sector with momentum $-nq$ is :
\begin{equation}
     \mathcal{V}_n(z) =
  \Bigl[\, i\lambda\cdot\partial X - n\,(q\cdot\psi)(\lambda\cdot\psi)\Bigr]
  e^{-in q\cdot X(z)} .
  \label{eq:Vn}
\end{equation}
Following the above notations this can be written as $\mathcal{V}_n=[B_1(\lambda)-nF_0(q)F_0(\lambda)]e^{-inq\cdot X}$.
The bosonic DDF operator with mode number $n\ge1$ is then defined as the zero mode of this vertex operator.

\begin{equation}
    \mathcal{A}_{-n}(\lambda) = \oint_0 \frac{\dd z}{2\pi i}\; \mathcal{V}_n(z)
\end{equation}
These operators satisfy the algebra of the oscillator modes
\begin{align}
    &[A_{-n}(\lambda),A_{-m}(\lambda)]=m(\lambda.\lambda)\delta_{n+m,0}=0\\
    \\
    &[A_{-n}(\lambda),A_{-m}(\lambda^{*})]=m(\lambda.\lambda^{*})\delta_{n+m,0}=m\delta_{m+n,0}
\end{align}
Let $|p_0,\nu\rangle$ denote the normalized massless state in the NS-NS sector of the closed superstring. This state has momentum $p_0$ and polarization $\nu$ in both chiral halves. This state is chosen to be in a particular quantum state satisfying :
$$\nu.q=0\quad q.p_0=1\quad\nu.\lambda=0\quad \nu^2=0$$ 
 Hence the seed state in picture number (-1,-1) is :
\begin{equation}
|p_0,\nu\rangle=\lim_{z,\bar z\to 0}
e^{-\phi-\tilde{\phi}}
(\nu\cdot\psi)(\nu\cdot\tilde{\psi})
e^{ip_0\cdot X}
|0\rangle
\end{equation}
A general state under consideration can thus be obtained by the action of the bosonic DDF operators on the seed state . This can be written as :
\begin{equation}
|K;p,\lambda,\nu\rangle\equiv\left(
\prod_{i=1}^{l}
\mathcal{A}_{-n_i}(\lambda)
\right)\left(
\prod_{j=1}^{l^{'}}
\tilde{\mathcal{A}}_{-\tilde{n}_j}(\lambda)
\right)
|p_0,\nu\rangle
=
\mathcal{O}
\left(
{\lambda},\nu;
\psi^i,\tilde{\psi^i},\partial X^i,\bar{\partial}X^{i}
\right)
|p\rangle.
\label{eq:p0nu}
\end{equation}

\noindent This state has a momentum $p=p_0-Kq$ and mass $M^2=2K$ for some integer $K\geq0$ representing the bosonic level and satisfy $p.q=1$. A compact form for $\mathcal{O}$ can be provided by noting that the above kinematic constraints results in the vanishing of almost all the contractions except between the exponential.

\noindent Then the holomorphic part of the operator $\mathcal{O}$ is given by:
\begin{equation}
  \;:\prod_{i=1}^{l}\ \!\Biggl[\;
     \sum_{d=1}^{n_i} \frac{B_d(\lambda)}{(d-1)!}\;\cS^{(n_i)}_{n_i-d} 
     - n_i\!\!\sum_{\substack{a,b\ge0\\ a+b\le n_i-1}}\!\!
       \frac{F_a(q)\,F_b(\lambda)}{a!\,b!}\;\cS^{(n_i)}_{n_i-1-a-b}
  \Biggr]\,F_0(\nu)\;:\ \equiv P_{\pi} .
  \label{eq:residue}
\end{equation}
Here $S^{(n)}_k$ are the elementary Schur polynomials evaluted on the variables $B_s(q)$. Here $\pi=(n_1\geq n_2\geq \cdots \geq n_l)$ represents a particular partition of the integer K with length $l(\pi)$ which we write as:
$$\pi=(1^{k_1}2^{k_2}\cdots n^{k_n}\cdots)\qquad k_n\geq 0\qquad \sum_{n=1}k_{n}n=K \qquad \sum_{n=1}k_n=l(\pi).$$
The two terms in \eqref{eq:residue} originate from the contractions of the two terms of \eqref{eq:Vn} with the operators in \eqref{eq:p0nu}.

\subsection{The family of states}
At a given level $K$, the states $|\pi,\tilde{\pi}\rangle$, labelled by all
pairs of partitions $(\pi,\tilde{\pi})$ of $K$, form the basis of subspace of closed-string
states used in the matrix computation. Since $\pi$ and $\tilde{\pi}$ are
chosen independently, the family contains $P(K)^2$ states, where $P(K)$ is
the number of integer partitions of $K$. 
The oscillator algebra of the DDF operators gives the following orthogonality relation:
\begin{equation}
    \langle\pi^{'}\tilde{\pi}^{'}|\pi,\tilde{\pi}\rangle=\delta_{\pi\pi^{'}}\delta_{\tilde{\pi}\tilde{\pi}^{'}}\mathcal{N}_{\pi}\mathcal{N}_{\tilde{\pi}}\qquad \cN_\pi = \prod_{n\ge1} n^{k_n}\, k_n! 
\end{equation}

\noindent For an explicit choice,
\begin{equation}
    \lambda=\frac{\hat{e_2}+i\hat{e_3}}{\sqrt{2}}\qquad\nu=\frac{\hat{e_4}+i\hat{e_5}}{\sqrt{2}}
\end{equation}
the state $|\pi,\tilde{\pi}\rangle$ is a weight vector with the non-zero value of the spin component 
$$J_{23}=l(\pi)+l(\tilde{\pi})\qquad J_{45}=2.$$
\subsection{Zero-picture vertices}
The state $|\pi,\tilde\pi\rangle$ is constructed in the $(-1,-1)$ picture.
On the torus the total picture number has to be zero in each chiral sector.
For the even spin structures the natural choice is thus to take all vertex
operators in the $(0,0)$ picture. This is achieved by the action of picture changing operator, which  acts on the holomorphic matter part as $G_{-\frac12}$, and
similarly with $\tilde G_{-\frac12}$ on the antiholomorphic side. The action of $G_{-\frac12}$ on the building blocks is:
\begin{equation}
G_{-\frac12}\cdot B_d(v)=F_d(v),\qquad
G_{-\frac12}\cdot F_d(v)=B_{d+1}(v),\qquad
G_{-\frac12}\cdot e^{ip\cdot X}=F_0(p)\,e^{ip\cdot X},
\end{equation}
and on products it acts as a graded derivation,
\begin{equation}
G_{-\frac12}\cdot(XY)=(G_{-\frac12}\cdot X)\,Y+(-1)^{|X|}\,X\,(G_{-\frac12}\cdot Y),
\end{equation}
where $|X|$ is the fermion number of $X$. The action of
$\tilde G_{-\frac12}$ on $\tilde B_d(v)$, $\tilde F_d(v)$ and
$e^{ip\cdot X}$ is identical. The matter part of the zero-picture vertex operator is then:
\begin{equation}
G_{-\frac12}\tilde G_{-\frac12}\Big(P_\pi\,\tilde P_{\tilde\pi}\,e^{ip\cdot X}\Big)
=Q_\pi\,\tilde Q_{\tilde\pi}\,e^{ip\cdot X},
\qquad
Q_\pi=G_{-\frac12}(P_\pi)-P_\pi\,F_0(p),
\end{equation}
where the minus sign arises because $P_\pi$ contains an odd number of
fermions, and $\tilde Q_{\tilde\pi}$ is defined in the same way. The full vertex operator is:
\begin{equation}
V^{(0,0)}_{\pi,\tilde\pi} \equiv V_{\pi,\tilde\pi} =c\bar c\,Q_\pi\,\tilde Q_{\tilde\pi}\,e^{ip\cdot X}.
\label{eq:V0}
\end{equation}

\noindent The fixed vertex has the usual
$c\bar c$ factor and the BRST completion discussed in
Appendix~\ref{app:spurious-pco}. Appendix~\ref{sec:explicitV} gives
explicit polynomials at $K=1,2$.

\section{One-loop self-energy}
\label{sec:one-loop}

We now turn to the torus two-point function. We place the integrated
vertex for $|\alpha,\alpha'\rangle$ at $z$ and the conjugate vertex for
$|\gamma,\gamma'\rangle$ at the origin. The calculation amounts to finding
the spin-summed correlator and integrating it over the insertion position
and the modulus. Including the norms of the two external states, the
reduced amplitude is
\begin{equation}
 A_{\gamma\gamma',\alpha\alpha'}=
 \frac{\displaystyle\int_{\cF}\dd^2\tau\,\tau_2^{-5}
       \int_{T_\tau}\dd^2z\,\mathcal I_{\gamma\gamma',\alpha\alpha'}(z,\tau)}
 {\sqrt{\cN_\alpha\cN_\gamma\cN_{\alpha'}\cN_{\gamma'}}}.
 \label{eq:Adef}
\end{equation}
Here $\cF$ is the fundamental domain of the modular group, $T_\tau$ is the torus
with modular parameter $\tau$, and $\mathcal I_{\gamma\gamma',\alpha\alpha'}$ is the
correlation function of the two vertex operators, including the sum over spin
structures, i.e.,
\begin{equation}
\mathcal I_{\gamma\gamma',\alpha\alpha'}(z,\tau)
=
\sum_{s,\tilde s=2,3,4}
w_s(\tau)\,\overline{w_{\tilde s}(\tau)}
\left\langle
V_{\gamma\gamma'}^\dagger(0)\,
V_{\alpha\alpha'}(z,\bar z)
\right\rangle_{s,\tilde s;\tau},
\end{equation}
where the brackets denote the normalized state torus correlators at
fixed spin structures,
and the conjugate vertex includes the BPZ phases.
The chiral weights $w_s(\tau)$ incorporate the non-zero-mode determinants
and GSO phases,
\begin{equation}
w_s(\tau)=
\frac{C_s\,\theta_s(0|\tau)^4}
     {2\,\theta_1'(0|\tau)^4},
\qquad C_3=1,\qquad C_2=C_4=-1.
\label{spin-structure}
\end{equation}
The factor $\tau_2^{-5}$ in \eqref{eq:Adef} comes from the ten-dimensional
loop-momentum integral. The plane waves give the Koba--Nielsen factor
\begin{equation}
  \bigl\langle e^{ip\cdot X(z,\bz)}\,e^{-ip\cdot X(0)}\bigr\rangle
  = e^{p^2 G(z)} = e^{-2K\,G(z)}
  = |E(z)|^{4K}\, e^{-4\pi K\,\tau_2\, y^2} ,
  \label{eq:KN}
\end{equation}
where we used $p^2=-2K$,  and $(\Im z)^2/\tau_2=\tau_2y^2$. The polynomials $Q_\alpha$, $\widetilde Q_{\alpha'}$ in $V_{\alpha,\alpha'}$ \eqref{eq:V0}, and the conjugates
of $Q_\gamma$, $\widetilde Q_{\gamma'}$ are contracted with each other and with the plane waves by Wick's theorem, following the
rules discussed in the next subsection.

The different factors in \eqref{spin-structure} correspond to different boundary conditions of the fermions. The $NS$ fermions with periodic boundary condition on torus compactification contributes a factor of $\sqrt{\theta_4(0|\tau) / \eta(\tau)}$, and there are $8$ of these, resulting into a factor of $\theta^4_4(0|\tau)$. Similarly, $NS$ and $R$ fermions with antiperiodic boundary condition on the torus gives $\theta_3(0|\tau)^4$ and $\theta_2(0|\tau)^4$ respectively. On the other hand, the $R$ fermions with periodic boundary condition have zero modes. These zero mode can be occupied or unoccupied, corresponding to the fermion number $F=1$ or $F=0$ respectively. Hence for the trace over fermions, we have $\text{Tr} ((-1)^F q^{L_0-\frac{c}{24}} )=0$, hence this boundary condition for $R$ fermions do not contribute. The trace over $8$ spacetime bosons provides an additional factor of $\eta(\tau)^{-8}$. Together with the fermionic trace this gives a factor of $\eta(\tau)^{12} \sim \theta'_1(0|\tau)^4$ in the denominator. The values of $C_s$ are determined by the GSO projection.

\subsection{Torus contractions}

Consider a field at the point $z$ (first vertex operator, momentum $p$, vectors
$p,q,\lambda,\nu$) and a field at the origin (second vertex operator, momentum $-p$,
vectors $-p,-q,\lambda^*,\nu^*$). Let $v,w$ be polarization vectors. The torus
propagators give the contractions in spin structure $s$.
\begin{align}
  \text{boson, cross:}\qquad
  &\bigl\langle B_d(v)(z)\,B_e(w)(0)\bigr\rangle
   = -(-1)^e\,(v\cdot w)\,G_{d+e}(z) ,
  \label{eq:rule-bb}\\
  \text{boson, same vertex:}\qquad
  &\bigl\langle B_d(v)\,B_e(w)\bigr\rangle_{\rm same}
   = -(-1)^e\,(v\cdot w)\,R_{d+e} ,
  \label{eq:rule-bbs}\\
  \text{fermion, cross:}\qquad
  &\bigl\langle F_d(v)(z)\,F_e(w)(0)\bigr\rangle_s
   = (-1)^e\,(v\cdot w)\,S_{s,d+e}(z) ,
  \label{eq:rule-ff}\\
  \text{fermion, same vertex:}\qquad
  &\bigl\langle F_d(v)\,F_e(w)\bigr\rangle_{s,\rm same}
   = (-1)^e\,(v\cdot w)\,T_{s,d+e} ,
  \label{eq:rule-ffs}\\
  \text{left--right:}\qquad
  &\bigl\langle B_1(v)\;\bar B_1(w)\bigr\rangle = h\,(v\cdot w) .
  \label{eq:rule-mixed}
\end{align}
Here $\bar B_1(w) = i w\cdot\bar\partial X$ is a right-moving field at either
vertex. In addition, a bosonic field can contract with the two plane waves. A field
$B_d(v)$ at $z$ gives, after summing its contraction with $e^{-ip\cdot X(0)}$ and its
self-contraction with $e^{ip\cdot X(z)}$,
\begin{equation}
  B_d(v)(z)\;\to\;(v\cdot p)\,\bigl[G_d(z) - R_d\bigr] , \label{eq:Bplane}
\end{equation}
and a field $B_e(w)$ at the origin gives $-(w\cdot p)\,\bigl[(-1)^eG_e(z) - R_e\bigr]$.
Since $\lambda\cdot p=\nu\cdot p=0$, only the $q$-fields have such contractions.
Fermionic contractions come with the usual signs of the Pfaffian (the sign of the
permutation that brings contracted pairs next to each other).
These rules, along with our kinematic choices imply:
\begin{itemize}
\item A $\lambda$-field at $z$ can only contract with a $\lambda^*$-field at the origin,
of the same statistics (boson with boson, fermion with fermion), or, if it is
$B_1(\lambda)$, with the right-moving field $\bar B_1(\lambda^*)$ at the origin. The
same holds for $\nu$.
\item Fields $F_d(p)$, $F_d(q)$ contract among themselves, at the same vertex or across,
with the weights $p\cdot p=-2K$, $p\cdot q=1$ (same vertex) and $2K$, $-1$ (across);
$q\cdot q=0$ in both cases.
\item Fields $B_d(q)$ never contract with each other ($q\cdot q=0$) nor with $\lambda$ or
$\nu$ fields; they only contract with the plane waves, by \eqref{eq:Bplane}.
\item Left--right contractions involve only the first-derivative fields $B_1(\lambda)$,
$B_1(\nu)$ (and their conjugates). Higher derivatives do not have left--right
contractions because $\partial_z^d\partial_{\bar w}^{e}G$ vanishes for $d\ge2$ or
$e\ge 2$ away from coincident points, and $q$-fields have $q\cdot q=q\cdot\lambda=0$.
\end{itemize}

\subsubsection*{Organising the left--right contractions}

The left--right contraction \eqref{eq:rule-mixed} is a constant. This
lets us do the contractions in two steps. We first choose the fields
paired across the chiralities, then contract the remaining fields within
each chirality. There are four kinds of left--right pairs, $B_1(\lambda)$ at $z$ with $\bar B_1(\lambda^*)$ at the origin, $B_1(\lambda^*)$ at
the origin with $\bar B_1(\lambda)$ at $z$, and the same two with $\nu$. Let
$a=(a_\lambda,a_\nu)$ be the numbers of left-moving fields $B_1(\lambda),B_1(\nu)$ at
$z$ used in such pairs, and $b=(b_\lambda,b_\nu)$ the numbers of left-moving fields
$B_1(\lambda^*),B_1(\nu^*)$ at the origin used in such pairs. Then the right-moving side
uses $a$ fields at the origin and $b$ fields at $z$. Counting the ways to choose the
fields and to pair them gives
\begin{align}
  \begin{split}
  \mathcal I_{\gamma\gamma',\alpha\alpha'}(z,\tau)
  ={}&(-1)^{2K-\ell(\alpha)-\ell(\alpha')}|E(z)|^{4K}e^{-4\pi K\tau_2y^2}\\
  &\times\sum_{a,b}\frac{h^{|a|+|b|}}{a!\,b!}
  \mathcal B_{a,b}(\alpha,\gamma)\,
  \overline{\mathcal B_{b,a}(\alpha',\gamma')} .
  \end{split}
  \label{eq:Iblocks}
\end{align}
The \emph{chiral block} $\mathcal{B}_{a,b}(z)$ is defined by
\begin{equation}
  \mathcal{B}_{a,b}(\alpha,\gamma)
  = \sum_{s=2,3,4} w_s(\tau)\;
    \Bigl\langle\,
      \bigl(\partial_{\lambda}^{a_\lambda}\partial_{\nu}^{a_\nu} Q_\alpha\bigr)(z)\;
      \bigl(\partial_{\lambda^*}^{b_\lambda}\partial_{\nu^*}^{b_\nu} Q'_\gamma\bigr)(0)
    \Bigr\rangle_s ,
  \label{eq:block}
\end{equation}
where $Q'_\gamma$ is the conjugate polynomial at the origin, $\partial_\lambda$ means
differentiation of the polynomial with respect to the variable $B_1(\lambda)$ (this
removes one such field and multiplies by the number of such fields present). The
correlator $\langle\cdots\rangle_s$ contains all the contractions within the
left-moving sector, including those with the plane waves at fixed spin structure $s$. The right-moving block is built by the same chiral procedure, but with its
own partitions $(\alpha',\gamma')$ and reversed removal labels $(b,a)$.
It is not generally the conjugate of the \emph{same} block. Here
$a!=a_\lambda!a_\nu!$ and similarly for $b!$; these factorials correct ordered
choices. The displayed sign is the product of the two BPZ phases, $(-1)^{K-\ell(\alpha)}(-1)^{K-\ell(\alpha')}$.

Counting transverse fields gives the necessary selection rules
\[
 \ell(\alpha)-a_\lambda=\ell(\gamma)-b_\lambda,\qquad a_\nu=b_\nu.
\]
The right block obeys the same rule with $a,b$ interchanged. Adding the two
length equations gives conservation of total $J_{23}$.

\subsubsection*{What a chiral block looks like}

Each block $\mathcal{B}_{a,b}$ is a sum over spin structures of polynomials in the
functions $G_n$, $R_n$, $S_{s,n}$, $T_{s,n}$ with rational coefficients. Two facts
about these polynomials will be important.

\paragraph{Dependence on $y$ and $h$.} The variable
$y$ enters only through $G_1=L_1-2\pi i y$, and the variable $h$ only through
$G_2=L_2-h$ and $R_2=U_2-h$. A block is therefore a polynomial in $y$ and $h$ whose
coefficients are meromorphic functions of $z$:
\begin{equation}
  \mathcal{B}_{a,b}(z) = \sum_{j,l} y^j\,h^l\;\mathcal{B}_{a,b;j,l}(z) .
  \label{eq:blockpoly}
\end{equation}
The finite set of indices $\bigl((a,b),j,l\bigr)$ specifies the chiral
coefficient functions used in the theta expansion.

\paragraph{Number of Szeg\H{o} kernels.} For states in the NS sector, for every Wick term the number of
cross-contractions of fermions is even. The reason is that $Q_\pi$ contains an even
number of fermions; the fermions of the first vertex that are not contracted among
themselves (an even number is removed by same-vertex contractions) must all be
contracted with fermions of the second vertex. Hence every term contains an even
number of factors $S_{s,n}(z)$.

\subsection{Carrying out the spin-structure sum}

It is the even number of Szeg\H{o} kernels in each Wick term that makes
the spin sum manageable. The coincidence terms $T_{s,n}$ are already
polynomials in the Weierstrass invariants $g_2,g_3$ and the half-period
values $e_s$ defined in Appendix~\ref{szego-derivative}. The
cross-contractions $S_{s,n}$ also reduce to polynomials when taken in
pairs, as the following lemma shows.

\begin{lemma}\label{lem:evenproducts}
An even product of the functions $S_{s,n}(z)$ is a polynomial in
$\wp(z)$ and $\wp'(z)$ with coefficients polynomial in $e_s$, $g_2$ and $g_3$.
\end{lemma}

The proof in Appendix~\ref{szego-derivative} uses
\begin{equation}
 (\wp')^2=4\wp^3-g_2\wp-g_3.
 \label{wp:eq}
\end{equation}
For any Wick term, reduce the resulting polynomial in $e_s$ modulo
$4e_s^3-g_2e_s-g_3$ and write its remainder as $P_0+P_1e_s+P_2e_s^2$.
The spin sum then replaces the full spin-dependent product by $-P_2/2$,
since $\sum_s w_s=\sum_s w_s e_s=0$ and $\sum_s w_s e_s^2=-1/2$
(Appendix~\ref{esk-sum}). The dependence on $z$ comes from
$\wp,\wp'$ and the completed bosonic propagators $G_n$.
They are doubly periodic at the physical value $y=\Im z/\tau_2$.
Multiplication by $E(z)^{2K}$ gives the transformation law used below.

\subsection{Theta basis and the insertion integral}

After the spin sum, write the chiral factor in \eqref{eq:Iblocks} as
\begin{equation}
 \Phi_{a,b}(z,y)=E(z)^{2K}\sum_{j,l}y^jh^l\mathcal B_{a,b;j,l}(z).
\end{equation}
The question now is how to integrate the product of two such factors over
the torus. It turns out that at any fixed level each factor can be written as
a finite linear combination of functions that pair to give Gaussian moments.
The basis has to accommodate the powers of $y$ as well as the theta
transformation law.

\begin{definition}\label{def:theta}
For an integer $K\ge1$, a residue $r\in\mathbb{Z}/2K\mathbb{Z}$ and an integer $j\ge0$, the
\emph{generalised theta function} of level $2K$ is
\begin{equation}
  \Th^{(j)}_{K,r}(z,y\,|\,\tau)
  = \sum_{\ell\,\in\, r+2K\mathbb{Z}}
    \Bigl(y+\frac{\ell}{2K}\Bigr)^{j}\; q^{\ell^2/(4K)}\; e^{2\pi i\ell z} .
  \label{eq:gentheta}
\end{equation}
Here $y$ is treated as an independent variable \footnote{At $y=\Im z/\tau_2$, these sums are related to Landau-level
wavefunctions on a torus with $2K$ units of magnetic flux
\cite{Onofri2001,HamadaKobayashi2012}. In \eqref{eq:gentheta},
 the monomial gets replaced by the Hermite polynomial and the sum gets multiplied by $e^{-2\pi K\tau_2y^2}$. Up to normalization
and a gauge phase, this gives the $j$th Landau-level orbital with
label $r$ (cf.\ Ref.~\cite{HamadaKobayashi2012}, Eq.~(2.28)).
    }. For $j=0$ these are the usual level-$2K$ theta functions,
$\Th_{K,r}:=\Th^{(0)}_{K,r}$. For example
\begin{equation}
  \Th_{1,0}(z|\tau) = \theta_3(2z\,|\,2\tau) ,\qquad \Th_{1,1}(z|\tau) = \theta_2(2z\,|\,2\tau) .
\end{equation}
\end{definition}

The series converges absolutely because of the Gaussian factor $q^{\ell^2/(4K)}$.
The generalised theta functions have the same periodicity as $E(z)^{2K}$,

\begin{lemma}\label{lem:thetaperiod}
As identities in the independent variable $y$, 
\begin{align}
  \Th^{(j)}_{K,r}(z+1,y) &= \Th^{(j)}_{K,r}(z,y) ,
  \nonumber\\
  \Th^{(j)}_{K,r}(z+\tau,y+1) &= q^{-K}e^{-4\pi i K z}\;\Th^{(j)}_{K,r}(z,y).
  \label{eq:thetaperiod}
\end{align}
\end{lemma}
We show this in appendix \ref{app:theta-trafo}. With $D=\partial_z+4\pi iK y$ acting at fixed $y$,
\begin{equation}
  D\,\Th^{(j)}_{K,r} = 4\pi i K\;\Th^{(j+1)}_{K,r} ,
  \label{eq:Drule}
\end{equation}
because $\partial_z$ brings down $2\pi i\ell = 4\pi i K\bigl[(y+\tfrac{\ell}{2K})-y\bigr]$.

The expansion follows from the transformation law and regularity of the chiral factors.
     \begin{proposition}[transformation law]\label{prop:P1}
For every $a,b$, and separately for every Wick term contributing to $\Phi_{a,b}$, as
identities in $y$ and $h$:
\begin{equation}
  \Phi_{a,b}(z+1,y) = \Phi_{a,b}(z,y) ,\qquad
  \Phi_{a,b}(z+\tau,y+1) = q^{-K}e^{-4\pi iKz}\,\Phi_{a,b}(z,y) .
  \label{eq:P1}
\end{equation}
\end{proposition}
 \begin{proposition}[no poles]\label{prop:P2}
For every $a,b$, every power $y^jh^l$, and separately for every Wick term, the
spin-structure sum $E(z)^{2K}\mathcal{B}_{a,b;j,l}(z)$ is an entire function of $z$.
More precisely, near $z=0$,
\begin{equation}
  E(z)^{2K}\,\mathcal{B}_{a,b;j,l}(z) = O\bigl(z^{\,2+|a|+|b|}\bigr) .
  \label{eq:P2}
\end{equation}
\end{proposition}

Appendix~\ref{app:prop_proof} proves the two propositions. Since $y$
enters through $G_1$ and $h$ through $G_2$ and $R_2$, the degrees are bounded
by $J_{a,b}\leq 2K+2-|a|-|b|$ and
$L_{a,b}\leq\lfloor(2K+2-|a|-|b|)/2\rfloor$.
The decomposition is
\begin{equation}
  \Phi_{a,b}(z,y)=\sum_{l=0}^{L_{a,b}} h^l\sum_{r=-K+1}^{K}\sum_{j=0}^{J_{a,b}}
  a^{(a,b)}_{r,j,l}(\tau)\;\Th^{(j)}_{K,r}(z,y|\tau).
  \label{eq:decomp}
\end{equation}
Here the coefficients $a_{r,j,l}^{(a,b)}(\tau)$ are obtained uniquely by matching expansion coefficients:
\begin{equation}
  \sum_{l,j\geq 0}^{L_{a,b}, J_{a,b}}h^l a^{(a,b)}_{r,j,l}(\tau)\,u^j
  =q^{-r^2/(4K)}\, \hat\Phi_{a,b;r}\Bigr(u - \frac{r}{2K}\Bigr),
  \quad -K<r\le K,
  \label{eq:coeffformula}
\end{equation}
In the above the Fourier coefficient is simply defined via:
\begin{equation}
    \hat{\Phi}_{a,b;r}(y) = \int_0^1 dx\, e^{-2\pi i\, r (x+i \,c)} \Phi_{a,b}\Bigr(x+ i\,c , y\Bigr).
\end{equation}
Appendix~\ref{app:decomp} derives the coefficient formula. At fixed
noncoincident $z$, the normalized prime form, propagators and even-spin
weights have finite cusp expansions. The sum over spin structures is
invariant under $\tau\mapsto\tau+1$ at fixed $z,y,h$, so the Fourier
coefficients of $\Phi_{a,b}$ have nonnegative integer powers of $q$.
It follows that
\begin{equation}
  a^{(a,b)}_{r,j,l}(\tau) = q^{-r^2/(4K)}\sum_{n\ge0} c^{(a,b)}_{r,j,l,n}\;q^{n},
  \qquad -K<r\le K.
  \label{eq:aexpansion}
\end{equation}
Once the coefficients have been obtained, by the procedure in
Appendix~\ref{app:numdet}, the integral over $z$ becomes particularly
simple. One torus cycle pairs equal Fourier modes, and the sum over
those modes unfolds the other cycle into a Gaussian integral on the
real line.

\begin{lemma}\label{lem:orth}
For residues $r,r'$ and integers $j,j'\ge0$,
\begin{equation}
  \int_{T_\tau}\dd^2z\;e^{-4\pi K\tau_2y^2}\;
  \Th^{(j)}_{K,r}(z,y)\;\overline{\Th^{(j')}_{K,r'}(z,y)}\;\Big|_{y=\Im z/\tau_2}
  = \delta_{r,r'}\;\tau_2\int_{-\infty}^{\infty}\dd u\;u^{\,j+j'}\,e^{-4\pi K\tau_2u^2} .
  \label{eq:orth}
\end{equation}
\end{lemma}
A short proof of the above is presented in appendix \ref{app:lemorth}.
The Gaussian moments are
\begin{equation}
  \mathsf{M}_k(\tau_2) := \int_{-\infty}^{\infty}\dd u\;u^{k}\,e^{-4\pi K\tau_2u^2}
  = \begin{cases}
      0 , & k\ \text{odd},\\[4pt]
      \dfrac{1}{\sqrt{4K\tau_2}}\;\dfrac{(k-1)!!}{(8\pi K\tau_2)^{k/2}} , & k\ \text{even},
    \end{cases}
  \label{eq:moments}
\end{equation}
Substituting \eqref{eq:decomp} and \eqref{eq:aexpansion} into
\eqref{eq:Iblocks}, orthogonality pairs only equal residues. The Gaussian
moments and the explicit factors of $\tau_2$ and $h$ give
\begin{equation}
  A_{\gamma\gamma',\alpha\alpha'} = \int_{\cF}\dd^2\tau\;\sum_{(m,\bar m,s)}\cC_{\gamma\gamma',\alpha\alpha'}(m,\bar m,s)\;
          \tau_2^{-s}\,q^{m}\,\bq^{\,\bar m} ,
  \label{eq:modint}
\end{equation}
where the sum runs over the labels
\begin{align}
  m &= n - \frac{r^2}{4K} ,\qquad \bar m = n' - \frac{r^2}{4K} ,
  \nonumber\\
  s &= \frac92 + |a|+|b| + l + l' + \frac{j+j'}{2} ,\qquad j+j'\ \text{even},
  \label{eq:labels}
\end{align}
and the coefficients are
\begin{equation}
 \begin{split}
 \cC_{\gamma\gamma',\alpha\alpha'}(m,\bar m,s)
 ={}&\sum\frac{(-1)^{2K-\ell(\alpha)-\ell(\alpha')}
 \pi^{|a|+|b|+l+l'}(j+j'-1)!!}
 {2\sqrt K\,(8\pi K)^{(j+j')/2}a!b!
 \sqrt{\cN_\alpha\cN_\gamma\cN_{\alpha'}\cN_{\gamma'}}}\\
 &\quad\times c^{(a,b)}_{r,j,l,n}(\alpha,\gamma)\,
 \overline{c^{(b,a)}_{r,j',l',n'}(\alpha',\gamma')}.
 \end{split}
 \label{eq:Ccoef}
\end{equation}
the sum being over all $(a,b,r,j,l,n,j',l',n')$ that produce the given labels
$(m,\bar m,s)$.  All allowed $a,b$ must be retained. Note that all sums other than $n,n'$ run over a finite range. The following properties are immediate.
\begin{itemize}
\item The difference $m-\bar m=n-n'$ is an integer, so each term is periodic under
$\tau_1\to\tau_1+1$.
\item Adjoint consistency relates the transposed external matrix entry:
$\cC_{\gamma\gamma',\alpha\alpha'}(\bar m,m,s)
=\overline{\cC_{\alpha\alpha',\gamma\gamma'}(m,\bar m,s)}$.
\item All the values of $s$ are half-integers, since $j+j'$ is even.
\item The smallest exponent is $m=-K/4$, from $r=K$ and $n=0$.
\end{itemize}
Appendix~\ref{app:S-theta} proves modular invariance before truncation.
We truncate the $n,n'$ series in the numerical calculation and test the
error by varying the retained order. On $\cF$,
$|q|\leq e^{-\pi\sqrt3}\simeq 0.0043$.

\subsection{Integral over the torus modular parameter}
\label{sec:tauint}

After the insertion integral, we evaluate the modular integral
\eqref{eq:modint}. We suppress external-state labels and write $A$ and
$\cC(m,\bar m,s)$ for each matrix element. These coefficients are assembled
from the coefficients $c_{r,j,l,n}$ through \eqref{eq:Ccoef}.
Numerically, these coefficients are obtained through successive Fourier transformations in $z$ and $\tau_1$, followed by a truncation of the corresponding $q$-series, the details of which are explained in Appendix~\ref{app:coefficientsector}, where we also check the numerical accuracy.

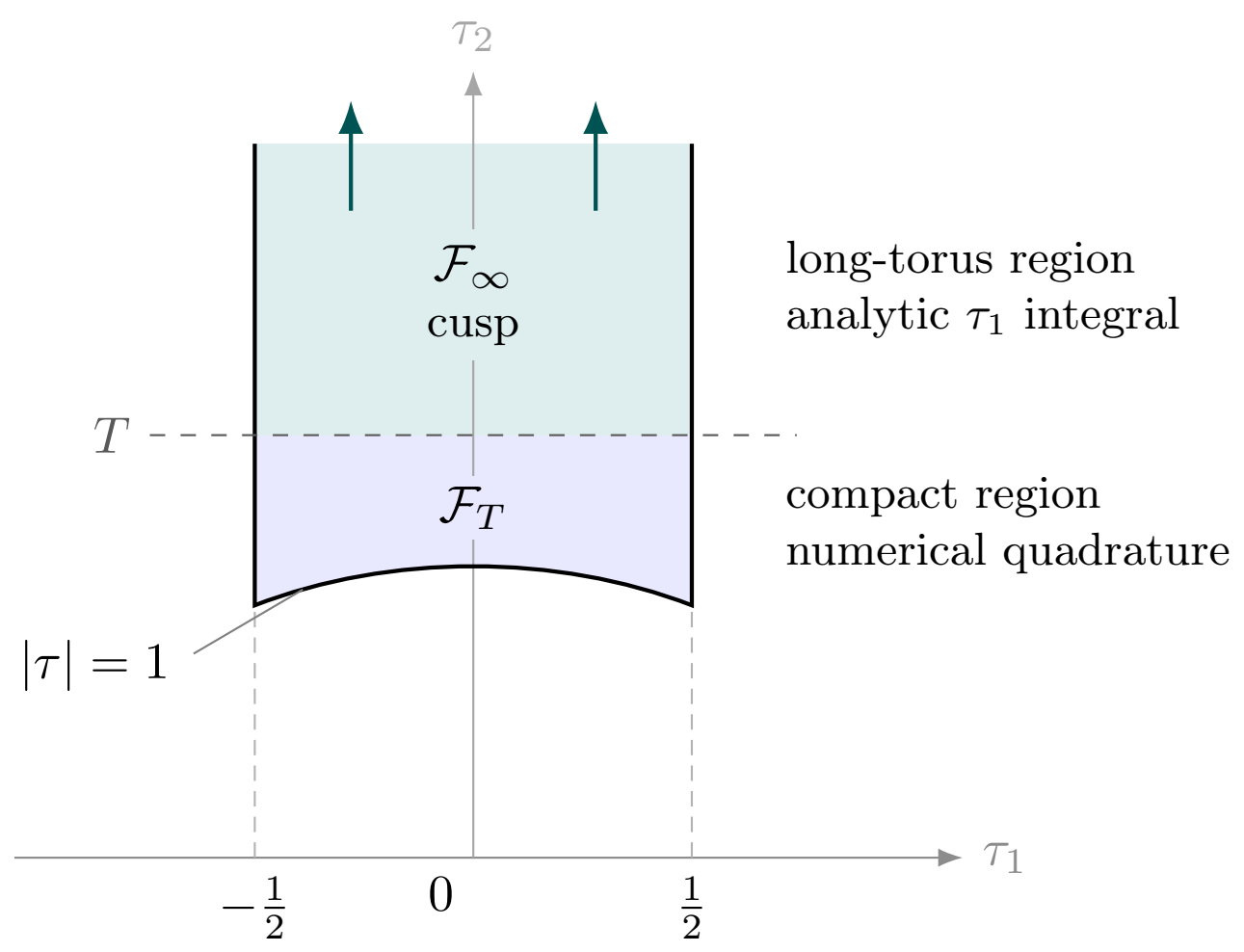
\begin{figure}[htbp]
\centering
\begin{tikzpicture}[x=3cm,y=2cm,>=Latex,font=\small]
  \fill[blue!9] (-.5,{sqrt(3)/2}) arc[start angle=120,end angle=60,radius=1]
    -- (.5,1.45) -- (-.5,1.45) -- cycle;
  \fill[teal!13] (-.5,1.45) rectangle (.5,2.45);
  \draw[->,black!45] (-1.05,0) -- (1.12,0) node[right] {$\tau_1$};
  \draw[->,black!35] (0,0) -- (0,2.70) node[above] {$\tau_2$};
  \draw[densely dashed,black!30] (-.5,0) -- (-.5,{sqrt(3)/2});
  \draw[densely dashed,black!30] (.5,0) -- (.5,{sqrt(3)/2});
  \draw[thick] (-.5,2.45) -- (-.5,{sqrt(3)/2})
    arc[start angle=120,end angle=60,radius=1] -- (.5,2.45);
  \draw[dashed,black!65] (-.74,1.45) node[left] {$T$} -- (.74,1.45);
  \node[fill=blue!9,inner sep=2pt] at (0,1.20) {$\cF_T$};
  \node[fill=teal!13,inner sep=3pt,align=center] at (0,1.93)
    {$\cF_\infty$\\[-1pt]\footnotesize cusp};
  \draw[->,teal!65!black,thick] (-.28,2.22) -- (-.28,2.60);
  \draw[->,teal!65!black,thick] (.28,2.22) -- (.28,2.60);
  \node[below] at (-.5,0) {$-\tfrac12$};
  \node[below] at (.5,0) {$\tfrac12$};
  \node[below left] at (0,0) {$0$};
  \node[anchor=east] at (-.65,.66) {$|\tau|=1$};
  \draw[black!50] (-.64,.70) -- (-.39,.92);
  \node[anchor=west,align=left,font=\footnotesize] at (.67,1.14)
    {compact region\\numerical quadrature};
  \node[anchor=west,align=left,font=\footnotesize] at (.67,1.95)
    {long-torus region\\analytic $\tau_1$ integral};
\end{tikzpicture}
\caption{The fundamental domain is split at a height $T\ge1$. Its curved
lower boundary is $\tau_2=\sqrt{1-\tau_1^2}$. The compact region $\cF_T$
is shaded blue and the cusp $\cF_\infty$ is shaded green; the latter extends
to $\tau_2=\infty$. The displayed height is schematic.}
\label{fig:modularregions}
\end{figure}

Using $q=e^{2\pi i\tau}$, a term in the integrand can be written as
\begin{equation}
  \tau_2^{-s}q^m\bq^{\,\bar m}
  =\tau_2^{-s}e^{2\pi i(m-\bar m)\tau_1}
    e^{-2\pi(m+\bar m)\tau_2}.
  \label{eq:taukernel}
\end{equation}
The distinction between the two parts of moduli space becomes apparent
in this form. In the interior all the retained Fourier pairs contribute,
while near the cusp the $\tau_1$ integral imposes level matching. The
surviving terms with $m=\bar m<0$, which we call polar terms, grow as
$\tau_2\to\infty$ and require analytic continuation. We can treat the two
regions separately by introducing a splitting height $T$.
Choose a splitting height $T\ge1$ and write
\begin{equation}
  \cF_T=\cF\cap\{\tau_2\le T\},\qquad
  \cF_\infty=\left\{\,|\tau_1|\le\frac12,\ \tau_2>T\,\right\}.
  \label{eq:tausplit}
\end{equation}
The two integration regions are shown in figure~\ref{fig:modularregions}.

\paragraph{The compact region.}

The contribution from $\cF_T$ is
\begin{equation}
\begin{split}
  A_{\rm compact}(T)
  ={}&\sum_{m,\bar m,s}\cC(m,\bar m,s)
  \int_{-1/2}^{1/2}\dd\tau_1
  \int_{\sqrt{1-\tau_1^2}}^T\dd\tau_2\,
  \tau_2^{-s}\\
  &\hspace{15mm}\times
  e^{2\pi i(m-\bar m)\tau_1}
  e^{-2\pi(m+\bar m)\tau_2}.
\end{split}
  \label{eq:Acompact}
\end{equation}
We use Gauss--Legendre quadrature in both variables. The curved lower
boundary prevents the $\tau_1$ integral from imposing $m=\bar m$;
all retained Fourier pairs contribute.

\paragraph{The cusp region.}
In $\cF_\infty$, the integration limits of the two variables are independent.
We first perform the $\tau_1$ integral. Since $m-\bar m=n-n'$ is an integer,
\begin{equation}
  \int_{-1/2}^{1/2}\dd\tau_1\,
  q^m\bq^{\,\bar m}
  =\delta_{m,\bar m}\,e^{-4\pi m\tau_2}.
  \label{eq:cuspmatching}
\end{equation}
Only level-matched Fourier terms survive; this does not restrict the
external-state labels. The remaining kernel is
\begin{equation}
  J_s(m;T)=\int_T^\infty\dd\tau_2\,
  \tau_2^{-s}e^{-4\pi m\tau_2}.
  \label{eq:Jsdef}
\end{equation}
For $m>0$, this converges exponentially. With $\tau_2=Tt$, it becomes
\begin{equation}
  J_s(m;T)=T^{1-s}E_s(4\pi mT),\qquad
  E_s(x)=\int_1^\infty\dd t\,t^{-s}e^{-xt},\quad \Re x>0.
  \label{eq:Jspositive}
\end{equation}
Here $E_s$ is the generalized exponential integral, distinct from the prime
form $E(z)$. For $m=0$, the integral also converges, since the values of $s$
in \eqref{eq:labels} satisfy $s\ge9/2$, and
\begin{equation}
  J_s(0;T)=\frac{T^{1-s}}{s-1}.
  \label{eq:Jszero}
\end{equation}

\begin{figure}[htbp]
\centering
\begin{tikzpicture}[x=1.8cm,y=1.65cm,>=Latex,font=\small]
  \draw[->,black!45] (-.15,0) -- (3.55,0) node[right] {$\Re t$};
  \draw[->,black!45] (0,-.20) -- (0,2.50) node[above] {$\Im t$};
  \draw[->,dashed,red!65!black,thick] (.8,0) -- (3.20,0);
  \node[below,align=center,text=red!65!black,font=\footnotesize] at (2.2,-.12)
    {real-axis tail\\divergent for $m<0$};
  \draw[->,teal!65!black,very thick] (.8,0) -- (.8,2.20);
  \node[above,text=teal!65!black] at (.8,2.20) {$T+i\infty$};
  \node[anchor=west,align=left] at (1.10,1.48)
    {$t=T+iu$, $u\ge0$\\[3pt]
     $\displaystyle e^{4\pi|m|t}=e^{4\pi|m|T}e^{4\pi i|m|u}$};
  \draw[->,black!60] (1.38,0) arc[start angle=0,end angle=90,radius=.58];
  \fill (.8,0) circle (1.8pt);
  \node[below left] at (.8,0) {$T$};
  \node[below left] at (0,0) {$0$};
\end{tikzpicture}
\caption{Deformed contour for the polar cusp integral after the $\tau_1$
integration. The variable $t$ is the complexified $\tau_2$. The upward
contour makes the exponential oscillatory and implements
$E_s(4\pi mT-i0)$.}
\label{fig:cuspcontour}
\end{figure}
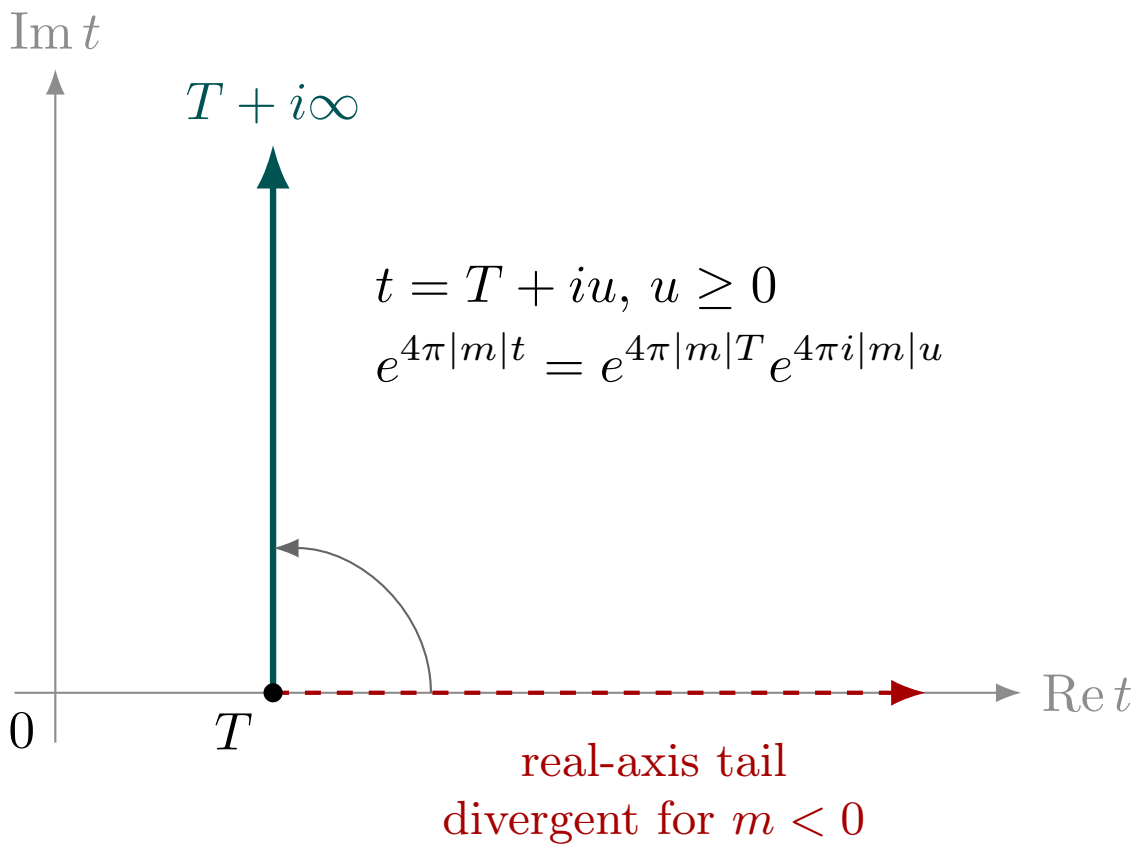

\paragraph{Analytic continuation of the growing terms.}
For the polar terms, the real-axis integral \eqref{eq:Jsdef} diverges. We define these
terms by analytic continuation, approaching the negative real axis of the
exponential-integral argument from below:
\begin{equation}
  J_s(m;T):=T^{1-s}E_s(4\pi mT-i0),\qquad m<0.
  \label{eq:Jscausal}
\end{equation}
Equivalently, the $i\varepsilon$ prescription rotates the tail into the
upper half-plane~\cite{Witten2013,Sen2016Contour},
\begin{equation}
  J_s(m;T)=\int_T^{T+i\infty}\dd t\,
  t^{-s}e^{4\pi|m|t},\qquad m<0.
  \label{eq:Jscontour}
\end{equation}
Figure~\ref{fig:cuspcontour} shows the contour used for these growing terms.
The exponential is oscillatory on this contour, and $t^{-s}$ ensures
convergence. For the half-integer values of $s$ in \eqref{eq:labels},
setting $X=4\pi|m|$ gives
\begin{equation}
  J_s(m;T)
  =-T^{1-s}\sum_{k=0}^\infty
    \frac{(XT)^k}{k!\,(k+1-s)}
    +i\,\frac{\pi X^{s-1}}{\Gamma(s)},\qquad m<0.
  \label{eq:Jsclosed}
\end{equation}
The imaginary part is independent of $T$, with its sign fixed by
\eqref{eq:Jscausal}. Numerically, we evaluate $J_s$ by the finite
error-function recurrence in appendix~\ref{app:validation_scope}, rather than truncating the series in \eqref{eq:Jsclosed}.

\paragraph{The integrated amplitude.}
Combining the compact region and the cusp gives
\begin{equation}
  A=A_{\rm compact}(T)
    +\sum_{m,s}\cC(m,m,s)J_s(m;T).
  \label{eq:Aintegrated}
\end{equation}
In our real basis, the compact contribution and the nonpolar cusp terms
give no imaginary part. Hence the absorptive matrix is determined entirely
by the polar terms,
\begin{equation}
  \Im A=\sum_{m<0}\sum_s\cC(m,m,s)
  \frac{\pi(4\pi|m|)^{s-1}}{\Gamma(s)}.
  \label{eq:Aabsorptive}
\end{equation}
In \eqref{eq:aexpansion}, $n\ge0$ labels the coefficient multiplying
$q^n$. Its prefactor $q^{-r^2/(4K)}$ gives the net power
$q^{n-r^2/(4K)}$, hence $m=n-r^2/(4K)$ as in \eqref{eq:labels}.
With theta residue $-K<r\le K$, a term is polar only if
$n<r^2/(4K)\le K/4$. Thus for $K=9,10$,
only $n=0,1,2$ can enter \eqref{eq:Aabsorptive}, subject to the
residue condition.

Finally, the splitting height $T$ is arbitrary. The continued cusp integral
satisfies the same endpoint relation as the convergent one,
\begin{equation}
  \frac{\partial J_s(m;T)}{\partial T}
  =-T^{-s}e^{-4\pi mT}.
  \label{eq:Jsderivative}
\end{equation}
The change of the compact contribution is the opposite boundary term:
\begin{equation}
  \frac{\dd A_{\rm compact}}{\dd T}
  =\sum_{m,s}\cC(m,m,s)T^{-s}e^{-4\pi mT}.
  \label{eq:compactboundary}
\end{equation}
Hence $\dd A/\dd T=0$. Varying $T$ checks the matching of the compact
and cusp contributions, independently of the coefficient checks in
appendix~\ref{app:coefficientsector}. The measured integration checks
are reported in appendix~\ref{app:validation_scope}. 

Using the modular properties of the $\tau, \bar{\tau}$ integrand for one loop string amplitudes, and several contour manipulations techniques \cite{BacciantiEtAl2025} developed a fast converging technology to perform the moduli integral. In appendix \ref{app:rademacher} we adapt the Rademacher method for our self-energy evaluations using crucially the modular properties of the $\Th_{K,r}^{(j)}$ coefficients, i.e. the $a_{r,j,l}^{(a,b)}(\tau)$ functions that appear in \eqref{eq:modint} (see appendix \ref{app:S-theta}) . 

\section{Results and conclusions}
\label{sec:results}

\subsection*{Matrix elements and low-level pole shifts}

We begin with the first few levels. At $K=1$ there is only one pair of
partitions in our family, and the calculation in
Appendix~\ref{app:K1-tauintegral} gives $A_{(1)}=0.002233377549032+i/210$.
The full level is one long Type II multiplet, so this one matrix element
is enough to find the common pole shift. At the next levels we have
several partitions to compare. Table~\ref{tab:diagonal-low} shows how
the diagonal entries vary for states $|\pi,\pi\rangle$ through $K=4$.

\begin{table}[htbp!]
\centering
\small
\renewcommand{\arraystretch}{1.15}
\begin{tabular}{c c r r c r r}
\toprule
$K$ & $\pi$ & $\Re A_\pi$ & $\Im A_\pi$ & variation & $\delta M_{\pi\pi}$ & $\Gamma_{\pi\pi}$ \\ \midrule
2 & $(2)$ & 0.0039412665 & 0.0065416065 & $1\cdot10^{-15}$ & $-0.009725$ & 0.032282 \\
2 & $(1,1)$ & 0.0036802234 & 0.0049783550 & $3\cdot10^{-16}$ & $-0.009081$ & 0.024567 \\
3 & $(3)$ & 0.0050256794 & 0.0083808553 & $4\cdot10^{-13}$ & $-0.010125$ & 0.033769 \\
3 & $(2,1)$ & 0.0042510802 & 0.0052901796 & $1\cdot10^{-15}$ & $-0.008564$ & 0.021315 \\
3 & $(1,1,1)$ & 0.0044102646 & 0.0055283503 & $3\cdot10^{-15}$ & $-0.008885$ & 0.022275 \\
4 & $(4)$ & 0.0057773203 & 0.0096058519 & $6\cdot10^{-12}$ & $-0.010080$ & 0.033519 \\
4 & $(3,1)$ & 0.0054318772 & 0.0077261862 & $3\cdot10^{-12}$ & $-0.009477$ & 0.026960 \\
4 & $(2,2)$ & 0.0080867172 & 0.0131829064 & $1\cdot10^{-13}$ & $-0.014109$ & 0.046001 \\
4 & $(2,1,1)$ & 0.0044554101 & 0.0050684167 & $1\cdot10^{-13}$ & $-0.007773$ & 0.017686 \\
4 & $(1,1,1,1)$ & 0.0049215572 & 0.0060146108 & $1\cdot10^{-13}$ & $-0.008587$ & 0.020988 \\
\bottomrule
\end{tabular}
\caption{Diagonal reduced amplitudes
$A_\pi=\langle\pi,\pi|A|\pi,\pi\rangle$. The variation is the
largest observed change under refinements of the Fourier grids and series
cutoff and changes of the split height $T$. The last two columns give the
corresponding diagonal mass-shift and width matrix elements at
$\alpha'=2$, $g_s=1$.}
\label{tab:diagonal-low}
\end{table}

At $K=2,3$ the known long-multiplet projectors give a linear relation
between the entries of the $\Sigma$ matrix for our choice of polarizations, and the physical complex
pole shifts. Table~\ref{tab:low-poles} gives the resulting mass shifts and
widths for all three long multiplets at $K=2$ and for the six multiplets
in the residual $\mathbf{44}\otimes\mathbf{44}$ at $K=3$.
Appendix~\ref{subsec:ddf_spin_degeneracy} gives the reconstruction.

\begin{table}[htbp]
\centering
\small
\begin{tabular}{clrr}
 \toprule
 $K$ & Superspin & $\delta M/g_s^2$ & $\Gamma/g_s^2$\\
 \midrule
 2 & $\mathbf1$ & $-0.01487747$ & $0.09399623$\\
 2 & $\mathbf{36}$ & $-0.00779239$ & $0.00913852$\\
 2 & $\mathbf{44}$ & $-0.00908059$ & $0.02456720$\\
 \midrule
 3 & $\mathbf1$   & $-0.02047261$ & $0.15354366$\\
 3 & $\mathbf{36}$  & $-0.01067618$ & $0.02045024$\\
 3 & $\mathbf{44}$  & $-0.01251194$ & $0.05388396$\\
 3 & $\mathbf{495}$ & $-0.00650712$ & $0.00797480$\\
 3 & $\mathbf{910}$ & $-0.00714851$ & $0.00989410$\\
 3 & $\mathbf{450}$ & $-0.00888503$ & $0.02227510$\\
 \bottomrule
\end{tabular}
\caption{Mass shifts and widths at $\alpha'=2$ obtained from the
low-level projector reconstruction. The $K=3$ entries are those
of $H\otimes H$ in \eqref{eq:residual-K3}.}
\label{tab:low-poles}
\end{table}

\FloatBarrier
\subsection{Charge and exchange blocks}

The $p(K)^2$ ordered pairs split into blocks labelled by $J_{23}$.
The two partition endpoints are $|(K),(K)\rangle$
at $J_{23}=2$ and $|(1^K),(1^K)\rangle$ at $J_{23}=2K$. 
Let $\mathsf P$ be the operator that
exchanges the left and right partitions,
$\mathsf P|\alpha,\alpha'\rangle=|\alpha',\alpha\rangle$. The two chiral halves of
the Type II string enter the calculation in the same way, and correspondingly the
computed matrix commutes with $\mathsf P$. Each block therefore splits into an even
and an odd part, spanned by
\begin{equation}
 |\alpha,\alpha'\rangle_\pm=\frac{|\alpha,\alpha'\rangle\pm|\alpha',\alpha\rangle}{\sqrt2}
 \quad(\alpha\ne\alpha'),
 \qquad
 |\alpha,\alpha\rangle_+=|\alpha,\alpha\rangle .
 \label{eq:exchangebasis}
\end{equation}
We refer to $(J_{23},\pm)$ as a sector. If $n_J$ is the dimension of
a charge block before resolving exchange, symmetry of $A$ leaves at most
$\sum_J n_J(n_J+1)/2$ independent entries before imposing exchange.
There are $4(K-1)$ nonempty charge and parity sectors for $K\ge2$.
Resolving the full $\mathrm{Spin}(9)$ and supersymmetry decomposition
requires the additional polarizations discussed in
Appendix~\ref{subsec:ddf_spin_degeneracy}.

\begin{figure}[ht]
\centering
\includegraphics[width=\textwidth]{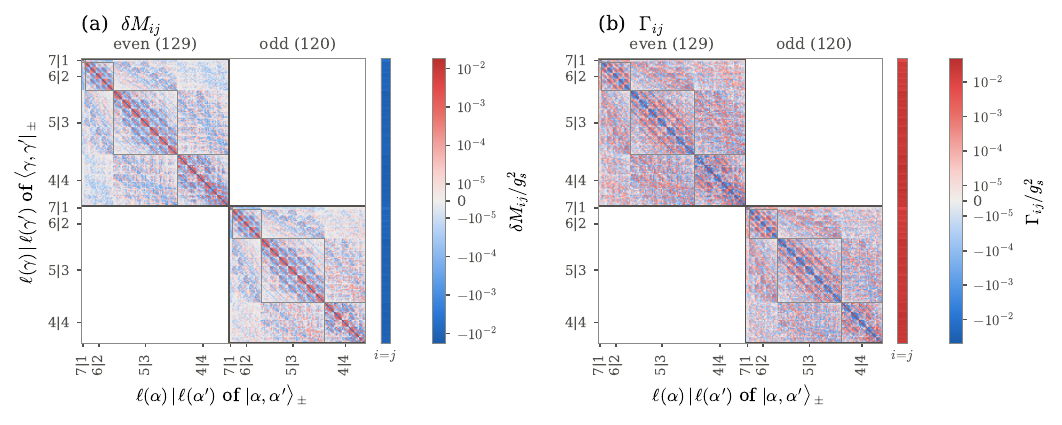}
\caption{Mass-shift matrix $D$ (a) and width matrix $W$ (b) at
$K=10$, $J_{23}=8$, with 129 exchange-even and 120 exchange-odd states.
Within each parity, the states are grouped by
$\ell(\alpha)|\ell(\alpha')$, with $\ell(\alpha)\ge\ell(\alpha')$.
The scale is linear for $|x|<10^{-5}$ and logarithmic in $|x|$ outside
that interval. Matrix entries are shown in units of $g_s^2$ with $\alpha'=2$.}
\label{fig:K10J8}
\end{figure}

\paragraph{A two-by-two example.}
At $K=9$, the exchange-odd sector with $J_{23}=16$ is spanned by
$|(1^9),(2^2,1^5)\rangle_-$ and $|(1^9),(3,1^6)\rangle_-$.
It gives a simple example of how two different oscillator partitions
can mix. The reduced amplitude in this basis is
\begin{equation}
 A=\begin{pmatrix}
 0.006068+0.005212\,i & -0.000648-0.000937\,i\\
 -0.000648-0.000937\,i & 0.006467+0.005866\,i
 \end{pmatrix}.
\end{equation}
The off-diagonal entry has both a real and an imaginary part, this is what results in a repulsion between the eigenvalues.
We can see the same two types of mixing on a larger scale at $K=10$,
$J_{23}=8$. Figure~\ref{fig:K10J8} shows the two real symmetric matrices
\begin{equation}
 D=\frac{\Re\Sigma}{2M_0},\qquad
 W=-\frac{\Im\Sigma}{M_0}.
 \label{eq:DW}
\end{equation}
The mass-shift and width matrices can be studied separately. When
they fail to commute, the real and imaginary parts of the complex pole
eigenvalues require diagonalizing the full complex matrix.

\subsection{Spacing statistics of the mass-shift matrix}
\label{sec:chaos}

The eigenvalues also give us a way to ask whether there is an
underlying random-matrix description. If there is, then the repulsion of neighbouring levels
should have characteristic spacing statistics. Since $D$ is real
symmetric, we compare to the Gaussian orthogonal ensemble (GOE)
\cite{BohigasGiannoniSchmit1984}. We arrange the eigenvalues $d_n$ in non-increasing order within
each charge and exchange sector and take ratios of neighbouring spacings,
$s_n=d_{n+1}-d_n$. The ratio $r_n=\min(s_n,s_{n+1})/\max(s_n,s_{n+1})$
can be used without unfolding. Its GOE surmise and Poisson law are
\cite{AtasEtAl2013}
\begin{equation}
 P_{\mathrm{GOE}}(r)=\frac{27}{4}\frac{r+r^2}{(1+r+r^2)^{5/2}},\qquad
 P_{\mathrm{P}}(r)=\frac{2}{(1+r)^2},\qquad 0\le r\le1.
 \label{eq:r-surmises}
\end{equation}
The corresponding means are about $0.536$ and $0.386$ for GOE and Poisson, respectively. When we include
all states at $K=10$, $J_{23}=8$, we find
$\langle r\rangle= 0.434357$, between the two reference values. At these
levels the same matrix can contain states built from rather different
partitions, from a single row or column to shapes near a staircase. Both these extremes are finely tuned. As we keep discarding these extremal shapes $\langle r \rangle$ value starts to increase. This suggests looking at the spacing distribution as we vary the Young tableaux shapes included in the calculation.

\paragraph{Typical partitions.}
To make this comparison we characterise the shape of a partition
$\alpha\vdash K$ by the number of standard Young tableaux,
given by the hook-length formula~\cite{FrameRobinsonThrall1954},
\begin{equation}
 f^\alpha=\frac{K!}{\prod_{b\in\alpha}h(b)}.
\end{equation}
Here $h(b)$ is the hook length at $b$. The number $f^\alpha$ is the
dimension of the corresponding representation of $S_K$. It counts the
ways in which one can build the Young diagram box by box. There is
only one such way for a row or a column, whereas the staircase
$(4,3,2,1)$ at $K=10$ admits $768$. For a pair of partitions we average
the logarithms of the two counts,
\begin{equation}
 C(\alpha,\alpha')=\tfrac12(\log f^\alpha+\log f^{\alpha'}).
\end{equation}
We rank $C$ over all $p(K)^2$ ordered pairs at level $K$, assigning
ties their midrank. We keep normalized ranks between 0.1 and 0.9.
At $K=10$ this window is
approximately $3.17\le C\le5.93$. We form the principal submatrix of $D$
on the retained states in each sector, diagonalize it, and compute spacing
ratios after discarding $10\%$ of the eigenvalues at each edge.
\begin{figure}[t]
\centering
\includegraphics[width=\textwidth]{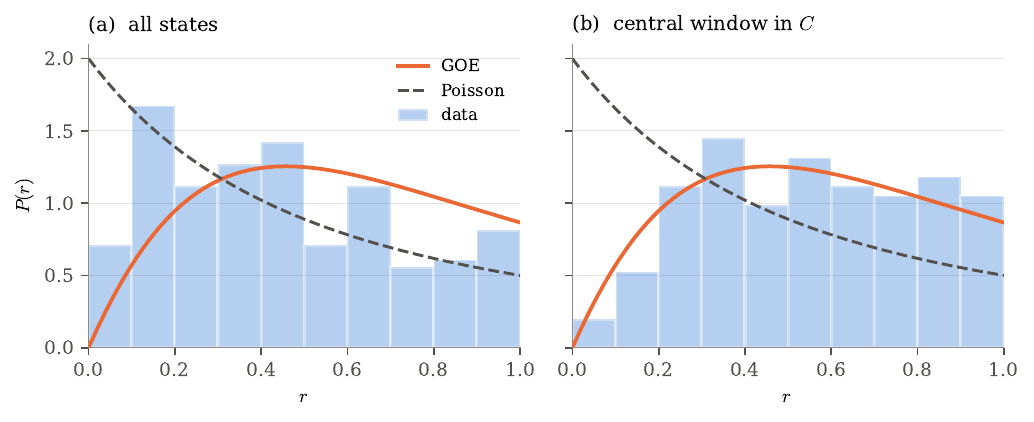}
\caption{Spacing ratios at $K=10$, $J_{23}=8$. Ratios are computed within
the two exchange parities, then pooled after trimming both spectral edges.
(a) All states (129 even and 120 odd, giving 197 ratios) give
$\langle r\rangle=0.449$. (b) The central $C$ window (99 even and 93 odd,
giving 152 ratios) gives $\langle r\rangle=0.559$.}
\label{fig:J8Pr}
\end{figure}

\begin{figure}[ht!]
\centering
\includegraphics[width=\textwidth]{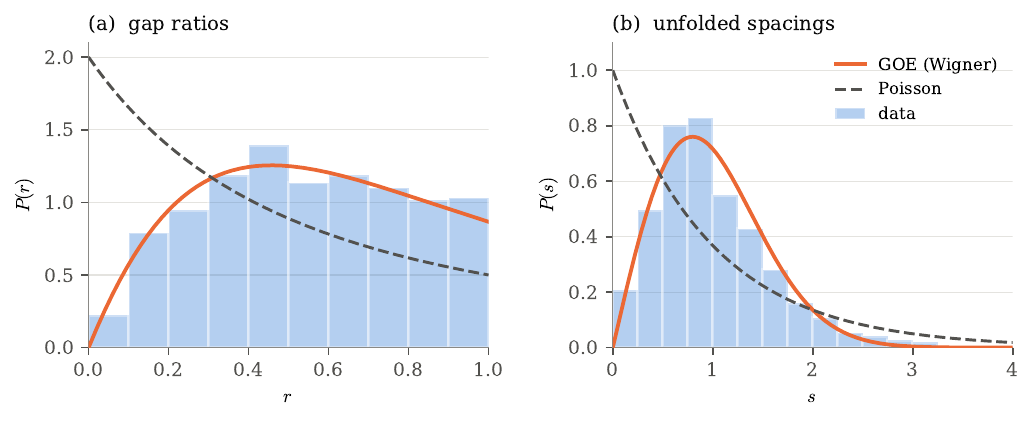}
\caption{Spacing statistics pooled over the two largest charge blocks
at each level: $J_{23}=7,6$ for $K=7$, $7,8$ for $K=8$, $8,7$ for $K=9$,
and $8,9$ for $K=10$. Each exchange block is restricted to the central
$C$ window and trimmed by $10\%$ at both spectral edges.
(a) Spacing ratios (582 ratios, $\langle r\rangle=0.549$).
(b) Unfolded spacings (598 spacings). For unfolding we use a
Gaussian-smoothed staircase in each block with bandwidth $\sigma n^{-1/5}$,
where $\sigma$ is the spectral standard deviation and $n$ the block size.
The spacings are scaled to unit mean.}
\label{fig:pooledRMT}
\end{figure}

The change is already visible in Fig.~\ref{fig:J8Pr}. With the central
window in $C$, there are fewer very small ratios and the distribution
is closer to the GOE curve. We can ask if the same pattern is visible
across several levels by pooling the two largest charge blocks at each
$K=7,\ldots,10$. The mean ratio is then $0.549$ in the central window
and $0.455$ for the full blocks. Figure~\ref{fig:pooledRMT} shows both
the ratios and the unfolded spacings for the selected partitions. 

\Needspace{9\baselineskip}
\subsection{Conclusions}
\label{sec:conclusions}

We have developed a method for carrying out the one-loop complex self-energy
computation for massive states in the DDF basis for superstrings in flat
$D=10$. We have focused on the bosonic partitions of excited states in
Type II. We have traded completeness of the basis for a view of how different
partitions can mix at a common mass, taking the computation up
to $K=10$. The ingredients of the method also give a way to ask the same
questions for other choices of physical states.

The main technical outcome of this work is the use of a generalised theta
basis that allows us to perform the worldsheet insertion integral exactly.
For a highly excited state at any finite level the required basis is still
finite. This is what makes the calculation tractable even when the vertex
polynomials come from different partitions. The absorptive matrix comes
from the long-torus region of the remaining modular integral. After analytic
continuation, only the polar terms that survive level matching contribute
to it.

At the first few levels we can see how the DDF matrix is related to the
physical poles. Supersymmetry gives a common shift at $K=1$. At $K=2$
we can recover all three complex shifts from the matrix elements, and
at $K=3$ we can do the same for six multiplets. At higher levels a full
little-group and supersymmetry analysis will require a larger choice of
external states. Meanwhile, what we have learnt from the mass-shift
matrices is that the shapes of the partitions matter for the spacing
statistics at the levels we can reach. We expect the distinction between
typical and atypical shapes to become sharper as the number of partitions
grows. It would be interesting to see whether the GOE-like behaviour in
the central window survives when the full symmetry sectors are available,
and how it is reflected in the complex poles.

\section*{Code and data availability}

The ancillary files accompanying this article on arXiv contain Python
code for the DDF torus kernels and for evaluating selected reduced
amplitudes. They include the direct \(K=1\) check of
Eq.~\eqref{eq:A1}, the Hermite-reduced \(K=9\) Rademacher checks in
Appendix~\ref{app:rademacher}, and the charge-resolved matrix blocks
and analysis code for the spacing statistics in Sec.~\ref{sec:chaos}.
The accompanying README gives the dependencies, example commands,
and reference outputs.

\subsection*{Acknowledgments} 

We are thankful to Sumit R. Das, Lorenz Eberhardt, Apratim Kaviraj, Ashoke Sen and Aninda Sinha for useful discussions. MV is also thankful to Raffaele Marotta and Ayush Tilloo for discussions and collaboration on related works. We also thank International Center for Theoretical Sciences (ICTS), Bengaluru for the hospitality during the program  ``Worldsheet String Theory and String Field Theory'' while the work was in its final stages. DD acknowledges the support from ANRF grant ANRF/ARG/2025/001338/PS and  the P. K. Kelkar Fellowship at IIT Kanpur. MV is also supported in part by the PRIUS (IITI/YFRSG-PRIUS/2023-24/Phase-I/03) component of the Young Faculty Research Seed Grant of IIT Indore.

\noindent\textbf{Statement on AI use.} The numerical coding and draft-preparation workflow made use of AI coding and reasoning assistants, including OpenAI Codex and ChatGPT and Anthropic's Claude, under human supervision. All physics assumptions, numerical choices, and conclusions are the responsibility of the authors.

\appendix
\addtocontents{toc}{\protect\setcounter{tocdepth}{1}}

\section{Low-level examples}
\label{sec:explicitV}

The first few partitions give short enough vertex polynomials to write
out explicitly. At $K=1$ we can follow the calculation all the way from
the contractions to the modular integral, and find the imaginary part
in closed form.

\subsection{Vertex polynomials at \texorpdfstring{$K=1,2$}{K=1,2}}
In the formulas below the fermions are written in the order
$\lambda$-type, $p$-type, $q$-type, $\nu$-type, the exponential $e^{ip\cdot X}$ is
implicit, and all products are normal ordered.

\paragraph{$K=1$, $\pi=(1)$.} The $(-1)$-picture polynomial is
\begin{equation}
  P_{(1)} = F_0(\lambda)F_0(q)F_0(\nu) + B_1(\lambda)F_0(\nu) ,
\end{equation}
and the zero-picture polynomial has seven terms,
\begin{align}
  Q_{(1)} ={}& -F_0(\lambda)F_0(p)F_0(q)F_0(\nu)
     \;+\; F_1(\lambda)F_0(\nu)
     \;+\; B_1(\lambda)F_0(p)F_0(\nu)
     \;+\; B_1(\lambda)F_0(q)F_0(\nu)
  \nonumber\\
   &+\; B_1(\lambda)B_1(\nu)
     \;-\; B_1(q)F_0(\lambda)F_0(\nu)
     \;+\; B_1(\nu)F_0(\lambda)F_0(q) .
  \label{eq:Q1}
\end{align}
We call these seven monomials $M_1,\dots,M_7$ in the order written.

\paragraph{$K=2$, $\pi=(1,1)$.} Here $\cN_{(1,1)}=2$ and
\begin{align}
  P_{(1,1)} ={}& 2B_1(\lambda)F_0(\lambda)F_0(q)F_0(\nu) + B_1(\lambda)^2F_0(\nu) ,
  \\
  Q_{(1,1)} ={}& -2F_0(\lambda)F_1(\lambda)F_0(q)F_0(\nu)
   - 2B_1(\lambda)F_0(\lambda)F_0(p)F_0(q)F_0(\nu)
   + 2B_1(\lambda)F_1(\lambda)F_0(\nu)
  \nonumber\\
  & + B_1(\lambda)^2F_0(p)F_0(\nu)
   + 2B_1(\lambda)^2F_0(q)F_0(\nu)
   + B_1(\lambda)^2B_1(\nu)
  \nonumber\\
  & - 2B_1(\lambda)B_1(q)F_0(\lambda)F_0(\nu)
   + 2B_1(\lambda)B_1(\nu)F_0(\lambda)F_0(q) .
\end{align}

\paragraph{$K=2$, $\pi=(2)$.} Here $\cN_{(2)}=2$. Using $\cS^{(2)}_1=-2B_1(q)$ in
\eqref{eq:residue},
\begin{align}
  P_{(2)} ={}& B_2(\lambda)F_0(\nu) - 2B_1(\lambda)B_1(q)F_0(\nu)
     + 2F_0(\lambda)F_1(q)F_0(\nu) + 2F_1(\lambda)F_0(q)F_0(\nu)
  \nonumber\\
   & - 4B_1(q)F_0(\lambda)F_0(q)F_0(\nu) ,
\end{align}
and the zero-picture polynomial has seventeen terms,
\begin{align}
  Q_{(2)} ={}& -2F_0(\lambda)F_0(p)F_1(q)F_0(\nu) - 4F_0(\lambda)F_0(q)F_1(q)F_0(\nu)
  \nonumber\\
  & - 2F_1(\lambda)F_0(p)F_0(q)F_0(\nu) + F_2(\lambda)F_0(\nu)
  \nonumber\\
  & - 2B_1(\lambda)B_1(q)F_0(p)F_0(\nu) - 4B_1(\lambda)B_1(q)F_0(q)F_0(\nu)
    - 2B_1(\lambda)B_1(q)B_1(\nu)
  \nonumber\\
  & + B_2(\lambda)F_0(p)F_0(\nu) + 2B_2(\lambda)F_0(q)F_0(\nu) + B_2(\lambda)B_1(\nu)
  \nonumber\\
  & + 4B_1(q)F_0(\lambda)F_0(p)F_0(q)F_0(\nu) - 4B_1(q)F_1(\lambda)F_0(\nu)
    + 4B_1(q)^2F_0(\lambda)F_0(\nu)
  \nonumber\\
  & - 4B_1(q)B_1(\nu)F_0(\lambda)F_0(q) - 2B_2(q)F_0(\lambda)F_0(\nu)
  \nonumber\\
  & + 2B_1(\nu)F_0(\lambda)F_1(q) + 2B_1(\nu)F_1(\lambda)F_0(q) .
\end{align}

\subsection{Contractions at \texorpdfstring{$K=1$}{K=1}}
\label{sec:K1contractions}

For $\pi=(1)$ we have $\cN_{(1)}=1$ and the seven monomials $M_1,\dots,M_7$ of
\eqref{eq:Q1}. The conjugate polynomial at the origin has the same seven monomials
with $(p,q,\lambda,\nu)\to(-p,-q,\lambda^*,\nu^*)$. Using the selection rules, a
monomial $M_i$ at $z$ and a monomial $M_j$ at the origin can only give a non-zero
contribution to $\mathcal{B}_{0,0}$ if their $\lambda$- and $\nu$-fields have the same
statistics. Since $T_{s,0}=0$ and $U_1=0$, the non-zero contributions are, writing
$S_n$ for $S_{s,n}(z)$:
\begin{center}
\renewcommand{\arraystretch}{1.25}
\begin{tabular}{lll}
\toprule
pair & contractions & contribution \\ \midrule
$M_1\times M_1$ & $\lambda\lambda^*,\ \nu\nu^*$, and $p$ with $-q$, $q$ with $-p$ & $-S_0^4$ \\
$M_2\times M_2$ & $F_1(\lambda)F_0(\lambda^*)$, $\nu\nu^*$ & $S_0S_2$ \\
$M_2\times M_6$, $M_6\times M_2$ & fermions $\lambda\lambda^*,\nu\nu^*$; $B_1(\mp q)$ with plane waves & $-G_1S_0S_1$ each\\
$M_3\times M_3$ & $B_1(\lambda)B_1(\lambda^*)$; $p$ with $-p$, $\nu\nu^*$ & $-2G_2S_0^2$ \\
$M_3\times M_4$, $M_4\times M_3$ & $B_1(\lambda)B_1(\lambda^*)$; $p$ with $-q$ (or $q$ with $-p$), $\nu\nu^*$ & $+G_2S_0^2$ each\\
$M_5\times M_5$ & $B_1(\lambda)B_1(\lambda^*)$, $B_1(\nu)B_1(\nu^*)$ & $G_2^2$ \\
$M_6\times M_6$ & $B_1(\pm q)$ with plane waves; $\lambda\lambda^*,\nu\nu^*$ & $G_1^2S_0^2$ \\
\bottomrule
\end{tabular}
\end{center}
As an example of how the entries arise, consider $M_6\times M_6$, i.e. the term
$-B_1(q)F_0(\lambda)F_0(\nu)$ at $z$ times $-B_1(-q)F_0(\lambda^*)F_0(\nu^*)$ at the origin.
The two bosons give $(G_1-R_1)(-G_1-R_1) = -G_1^2$ (since
$R_1=U_1=0$). The four fermions $F_0(\lambda)F_0(\nu)F_0(\lambda^*)F_0(\nu^*)$ are
contracted as $(13)(24)$, with sign $-1$, giving $-S_0^2$. The product with the
coefficient $(-1)(-1)$ is $G_1^2S_0^2$. Summing the table, the left-moving block
before the spin-structure sum is
\begin{equation}
  \mathcal{B}^{(s)}_{0,0}
  = G_1^2S_0^2 - 2G_1S_0S_1 + G_2^2 - S_0^4 + S_0S_2 .
  \label{eq:K1raw}
\end{equation}
The terms with $G_2S_0^2$ cancel already before the spin-structure sum. The only
other blocks are those with one left--right pair of each kind:
$\mathcal{B}_{(0,1),(0,1)}$ and $\mathcal{B}_{(1,0),(1,0)}$, both equal to
$\sum_s w_s\,G_2$, and $\mathcal{B}_{(1,1),(1,1)}=\sum_s w_s\cdot 1$. These three are
zero after the spin-structure sum, by Jacobi's identity.

\subsection{Spin-structure sum at \texorpdfstring{$K=1$}{K=1}}
\label{sec:K1spinsum}

The terms $G_1^2S_0^2$ and
$G_1S_0S_1$ vanish after the sum, the term $G_2^2$ vanishes by
Jacobi's identity, and the two remaining terms give
\begin{equation}
  \mathcal{B}_{0,0}
  = \sum_s w_s\bigl(S_s S_{s,2}-S_s^4\bigr)
  = \tfrac12 + \tfrac12 = 1 .
  \label{eq:K1block}
\end{equation}

Indeed, $S_s^2=\wp-e_s$ and
$S_sS_{s,1}=\wp'/2$, so their spin sums vanish. The quadratic
coefficients in $S_s^4=(\wp-e_s)^2$ and
$S_sS_{s,2}=(\wp-e_s)(2\wp+e_s)$ are $1$ and $-1$, respectively.
The rule in Appendix~\ref{esk-sum} then gives the two terms in
\eqref{eq:K1block}. Since the mixed blocks vanish, the integrand is
\begin{equation}
  \mathcal{I}_{(1)}(z,\tau) = |E(z)|^{4}\,e^{-4\pi\tau_2y^2} .
  \label{eq:K1integrand}
\end{equation}

\subsection{Insertion integral at \texorpdfstring{$K=1$}{K=1}}
\label{app:K1-zintegral}
In order to integrate \eqref{eq:K1integrand} we decompose into the $\Th^{(j)}_{K,r}$ functions. The relevant part is simply:
\begin{align}
    E(z)^2 &= \frac{\theta_2(0|2\tau)\,\Th_{1,0}(z|\tau) - \theta_3(0|2\tau)\,\Th_{1,1}(z|\tau)}{4\pi^2\,\eta(\tau)^6} .
  \label{eq:E2decomp}
\end{align}
Only the $j=j'=0$ term appears. The above can be done exactly by using the identities: $\theta_1(z|\tau)^2=\theta_2(0|2\tau)\,\theta_3(2z|2\tau)-\theta_3(0|2\tau)\,\theta_2(2z|2\tau)$
 and $\theta_1'(0)^2=4\pi^2\eta^6$. Thus the worldsheet integral using \eqref{eq:orth} and \eqref{eq:moments} give:
 \begin{align}
  \int_{T_\tau}\dd^2z\;|E(z)|^4e^{-4\pi\tau_2y^2}
  &= \frac{\tau_2\,\mathsf{M}_0(\tau_2)}{16\pi^4|\eta|^{12}}
     \Bigl(|\theta_2(0|2\tau)|^2+|\theta_3(0|2\tau)|^2\Bigr)
  \nonumber\\
  &= \frac{\sqrt{\tau_2}}{32\pi^4|\eta|^{12}}
     \Bigl(|\theta_2(0|2\tau)|^2+|\theta_3(0|2\tau)|^2\Bigr) ,
\end{align}
so that
\begin{equation}
  A_{(1)} = \int_{\cF}\dd^2\tau\;\frac{\tau_2^{-9/2}}{32\pi^4\,|\eta(\tau)|^{12}}
  \Bigl(|\theta_2(0|2\tau)|^2+|\theta_3(0|2\tau)|^2\Bigr) .
  \label{eq:A1}
\end{equation}
The $q$-expansions are $\theta_3(0|2\tau)=1+2q+2q^4+\dots$,
$\theta_2(0|2\tau)=2q^{1/4}(1+q^2+q^6+\dots)$ and
$\eta^{-6}=q^{-1/4}(1+6q+27q^2+\dots)$. Hence the term with $\theta_3$ starts with
$|q|^{-1/2}=q^{-1/4}\bq^{-1/4}$, that is $m=\bar m=-\tfrac14=-K/4$ (two massless decay
products), and the term with $\theta_2$ starts with $m=\bar m=0$.

\subsection{Modular integral at \texorpdfstring{$K=1$}{K=1}}
\label{app:K1-tauintegral}

The only level-matched polar term has $m=\bar m=-\tfrac14$ and comes from the leading term
of $|\theta_3(0|2\tau)|^2/|\eta|^{12}=|q|^{-1/2}(1+\dots)$, with $s=\tfrac92$ and
$\cC=\tfrac{1}{32\pi^4}$. By \eqref{eq:Aabsorptive}, with $4\pi|m|=\pi$ and
$\Gamma(\tfrac92)=\tfrac{105}{16}\sqrt\pi$,
\begin{equation}
  \Im A_{(1)} = \frac{1}{32\pi^4}\cdot\frac{\pi\cdot\pi^{7/2}}{\Gamma(\frac92)}
  = \frac{1}{32\pi^4}\cdot\frac{16\,\pi^4}{105} = \frac{1}{210} .
\end{equation}
The real part is the sum of the compact part and the real parts of the cusp
terms. With $T=1$ and $q$-expansions to order $q^{39}$, an independent evaluation of
\eqref{eq:A1} gives
\begin{align}
  \text{compact part} &= 0.000399750853220 ,\nonumber\\
  \text{cusp, real part} &= 0.001833626695812 ,\nonumber\\
  A_{(1)} &= 0.002233377549032 + \frac{i}{210} ,
\end{align}
with $\cN=\pi^2$ ($\alpha'=2$, $g_s=1$) and $M=\sqrt2$: $\delta M_{(1)}=-0.007793$ and
$\Gamma_{(1)}=\pi^2/(210\sqrt2)=0.033233$.

\section{Spin sums and theta identities}
\label{proofs}

\subsection{Derivatives of the Szeg\H{o} kernel}
\label{szego-derivative}

The even Szeg\H{o} kernels obey $S_s^2=\wp-e_s$, where
\begin{align}
 e_2&=\wp(\tfrac12)=\tfrac{\pi^2}{3}(\theta_3^4+\theta_4^4),&
 e_3&=\wp(\tfrac{1+\tau}{2})=\tfrac{\pi^2}{3}(\theta_2^4-\theta_4^4),\nonumber\\
 e_4&=\wp(\tfrac{\tau}{2})=-\tfrac{\pi^2}{3}(\theta_2^4+\theta_3^4).
 \label{eq:es}
\end{align}
The theta constants are evaluated at zero. The $e_s$ are the three
roots of $4x^3-g_2x-g_3$. Differentiating $S_s^2$ and using
\eqref{wp:eq} gives
\begin{equation}
 S_s S_s'=\tfrac12\wp',\qquad
 (S_s')^2=\wp^2+e_s\wp+e_s^2-\tfrac14 g_2,\qquad
 S_s''=(2\wp+e_s)S_s.
 \label{del-Ss}
\end{equation}

\begin{proof}[Proof of Lemma~\ref{lem:evenproducts}]
Write $S_{s,n}=A_n S_s+B_n S_s'$, starting from $(A_0,B_0)=(1,0)$.
Differentiation gives
$A_{n+1}=\partial_z A_n+(2\wp+e_s)B_n$ and
$B_{n+1}=A_n+\partial_z B_n$. The differential equation for $\wp$ \eqref{wp:eq},
keeps these coefficients polynomial in $\wp,\wp',e_s,g_2,g_3$.
In an even product the factors $S_s,S_s'$ can be paired. Each pair
is polynomial by $S_s^2=\wp-e_s$ and \eqref{del-Ss}. The same
differential equation fixes the Laurent expansion at $z=0$ and hence
the regular coefficients $T_{s,n}$.
\end{proof}

\subsection{Even-spin-structure weights}
\label{esk-sum}

Set $\mu_k=\sum_s w_s e_s^k$. Substitution of \eqref{eq:es} and the
theta identities $\theta_3^4=\theta_2^4+\theta_4^4$ and
$\theta_1'(0)=\pi\theta_2\theta_3\theta_4$ gives
\begin{equation}
 \mu_0=\mu_1=0,\qquad \mu_2=-\tfrac12,\qquad
 \mu_{k+3}=\tfrac14(g_2\mu_{k+1}+g_3\mu_k).
 \label{mu-recursion}
\end{equation}
The recurrence follows from the cubic for $e_s$. Equivalently, divide a
polynomial $f(x)$ by $4x^3-g_2x-g_3$ and write its remainder, which is at most quadratic, as
$\alpha+\beta x+\gamma x^2$. Then $\sum_s w_s f(e_s)=-\gamma/2$.

\subsection{Theta translation law}
\label{app:theta-trafo}
\begin{proof}
The first relation holds because $\ell$ is an integer. For the second, the term with
label $\ell$ becomes
$(y+1+\tfrac{\ell}{2K})^j q^{\ell^2/(4K)+\ell}e^{2\pi i\ell z}$. Since
$\tfrac{\ell^2}{4K}+\ell = \tfrac{(\ell+2K)^2}{4K}-K$ and
$e^{2\pi i\ell z}=e^{2\pi i(\ell+2K)z}e^{-4\pi iKz}$, this is
$q^{-K}e^{-4\pi iKz}$ times the term with label $\ell+2K$ of the original series.
Summing over $\ell$ gives the claim. For $E^{2K}$, use
$\theta_1(z+1)=-\theta_1(z)$ and $\theta_1(z+\tau)=-q^{-1/2}e^{-2\pi iz}\theta_1(z)$, raised to
the power $2K$.
\end{proof}

\subsection{Chiral translation law and regularity}
\label{app:prop_proof}

\begin{proof}[Proof of Proposition~\ref{prop:P1}]
Check the translation of each contraction function and of $E^{2K}$.
\begin{itemize}
\item $L_n$ with $n\ge2$ is a derivative of $\log\theta_1$ of order at least two, hence
doubly periodic. $L_1=-\partial\log\theta_1$ satisfies $L_1(z+1)=L_1(z)$ and
$L_1(z+\tau)=L_1(z)+2\pi i$, because $\log\theta_1(z+\tau)=\log\theta_1(z)-2\pi iz+\text{const}$.
Hence $G_1=L_1-2\pi iy$ is unchanged under $(z,y)\to(z+\tau,y+1)$, and also under
$z\to z+1$. The functions $G_n$, $n\ge2$, are unchanged as well.
\item $R_n$, $U_n$, $T_{s,n}$ and $h$ do not depend on $z$.
\item By the periodicities of the theta functions,
$S_s(z+1)=\epsilon_sS_s(z)$ and $S_s(z+\tau)=\epsilon'_sS_s(z)$, with
$(\epsilon_2,\epsilon'_2)=(+,-)$, $(\epsilon_3,\epsilon'_3)=(-,-)$,
$(\epsilon_4,\epsilon'_4)=(-,+)$. The derivatives $S_{s,n}$ have the same signs. Every
Wick term contains an even number of factors $S_{s,n}$, so the
signs cancel.
\item $E^{2K}$ supplies the factor $q^{-K}e^{-4\pi iKz}$ (lemma~\ref{lem:thetaperiod}).
\end{itemize}
\end{proof}

\begin{proof}[Proof of Proposition~\ref{prop:P2}]
For this proof we need the near zero behavior of the Szeg\H{o} kernel and it's derivatives.
\begin{align}
    &S_s(z) \sim \frac{1}{z} \left( 1 - \frac{e_s}{2}z^2 + \left( \frac{g_2}{40}-\frac{e_s^2}{8} \right) z^4 + \mathcal{O}(z^6) \right), \nonumber \\
    &S'_s(z) \sim -\frac{1}{z^2} \left( 1 + \frac{e_s}{2}z^2 - 3 \left( \frac{g_2}{40}-\frac{e_s^2}{8} \right) z^4 + \mathcal{O}(z^6) \right), \nonumber \\
    &S_{s,n}(z) \sim \frac{(-1)^n n!}{z^n} \left( 1 + \mathcal{O}(z^{2 \lceil \frac{n+1}{2} \rceil}) \right).
\end{align}
For each $n$-th order derivative of $S_s$ the order $z^{1-n}$ term is present iff $n \leq 1$. For all the higher derivatives, this term is not present. Using above Laurent series expansions any even product of the Szeg\H{o} kernels can be written as following.
\begin{align}
    \prod_{i=1}^{2k} S_{s,n_i}(z) \sim \prod_i  \left( \frac{ (-1)^{n_i} n_i!}{z^{1+n_i}} \right) \left( 1 + \frac{A_1-A_0}{2} e_s z^2 + \mathcal{O}(z^4)\right).
\end{align}
In the above expression $A_0$ is the number of derivative orders $n_i=0$ and $A_1$ is the number of $n_i=1$ in the even product. The $\mathcal{O}(z^2)$ term inside the bracket comes with this specific factor which is proportional to $e_s$ only. But the $\mathcal{O}(z^4)$ term inside the bracket can have a coefficient $e_s^2$. This way after doing the spin sum the first $2$ terms give zero contribution \eqref{mu-recursion}. The minimum non-vanishing contribution comes from the second subleading term of the expansion.
\begin{align}
    \sum_s w_s \prod_{i=1}^{2k} S_{s,n_i}(z) \sim z^{4 - \sum_{i=1}^{2k}(1+n_i) }.
    \label{szego:spin-sum}
\end{align}
Each term of the block $\mathcal{B}_{a,b}$ is aspin sum of objects, which are product of derivatives of bosonic correlators $G_n$ and even product of Szeg\H{o} kernels. Near $z \to 0$, $G_n$ has a pole of order $n$.
\begin{align}
    \prod_{i=1}^{2k}S_{s,n_i} \prod_{j} G_{l_j} \sim \frac{1}{z^{W}}; \qquad \text{where, } W = \sum_i (n_i+1) +\sum_j l_j = 2K+2 - |a| - |b|.
    \label{block-product}
\end{align}
Using \eqref{szego:spin-sum} and \eqref{block-product} we can derive the $z \to 0$ behavior of the block $\mathcal{B}_{a,b}$. The Koba--Nielsen factor is regular in the $z \to 0$ limit, i.e.$E^{2K} \sim z^{2K}$. 
\begin{align}
    & \mathcal{B}_{a,b} \sim \sum_s \prod_{i=1}^{2k}S_{s,n_i} \prod_{j} G_{l_j} \sim z^{4-W} \nonumber \\
    & E^{2K} \mathcal{B}_{a,b} \sim z^{2 + |a| +|b|}.
    \label{B_ab-structure}
\end{align}
Hence, $E^{2K} \mathcal{B}_{a,b}$ has no singularity at $z \to 0$, it is $O(z^{2K+4-W})= O(z^{2+|a|+|b|})$. As $|a|, |b| \geq 0$, this term is always regular. Since, $y$ and $h$ does not have any further singularity in $z \to 0$, this argument also holds true for the coefficients $E^{2K} \mathcal{B}_{a,b;j,l}(z)$. 
\end{proof}

\subsection{Proof of the decomposition coefficient formula \eqref{eq:decomp}}
\label{app:decomp}
We want to show that given a function with the (quasi)periodic properties:
\begin{align}
F(z+1,y) &= F(z,y), \quad F(z+\tau,y+1) = q^{-K} e^{-4\pi i K z } F(z,y)
\end{align}
and regular $z \rightarrow 0$ behaviour of coefficients in a $y^{m}, h^n $ finite polynomial expansion, the following decomposition holds:
\begin{equation}
  F(z,y)=\sum_{l=0}^{L} h^l\sum_{r=-K+1}^{K}\sum_{j=0}^{J}
  a_{r,j,l}(\tau)\;\Th^{(j)}_{K,r}(z,y|\tau),
  \label{prop:gen}
\end{equation}
where the coefficients $a_{r,j,l}^{}(\tau)$ are obtained from expanding:
\begin{equation}
  \hat F_{r}\Bigr(u - \frac{r}{2K}\Bigr) = q^{r^2/(4K)} \sum_{l,j\geq 0}^{L, J}h^l a^{}_{r,j,l}(\tau)\,u^j
  \, , 
  \quad -K<r\le K.
  \label{eq:coeffformula1}
\end{equation}
In the above the Fourier coefficient is simply defined via:
\begin{equation}
    \hat F_{r}(y) = \int_0^1 dx\, e^{-2\pi i\, r (x+i \,c)} F\Bigr(x+ i\,c , y\Bigr).
\end{equation}
Equation \eqref{eq:coeffformula1} is the defining relationship for $a_{r,j,l}$.
The $\Th^{(j)}_{K,r}$ are the generalized theta functions of level $2K$ as defined in \eqref{eq:gentheta}.

\begin{proof}
We start by writing the Fourier series that first property of \eqref{prop:gen} allows us: 
\begin{align}
    F(z,y) &= \sum_{l'=-\infty}^\infty \hat{F}_{l'}(y) e^{2\pi i l'z}  = \sum_{l = 0}^{L} h^l  \sum_{l'=-\infty}^\infty \sum_{j = 0}^J  a_{l',j,l}(\tau) \,e^{2\pi i l'z} q^{l'^2/(4K)}  \Bigr( y + \frac{l'}{2K} \Bigr)^j.\label{eq:step1}
\end{align}
In the second equality we used \eqref{eq:coeffformula1} and pulled a finite summation outside. Note that 
\begin{equation}
  \Th^{(j)}_{K,r}(z,y\,|\,\tau)
  = \sum_{\ell\,\in\, r+2K\mathbb{Z}}
    \Bigl(y+\frac{\ell}{2K}\Bigr)^{j}\; q^{\ell^2/(4K)}\; e^{2\pi i\ell z} .
  \label{eq:gentheta1}
\end{equation}
Now notice, in order to bring \eqref{eq:step1} to the form of \eqref{prop:gen}, we may first organize $l'$ sum as: 
$$
\sum_{l' = -\infty}^\infty \, f(l')= \sum_{r = -K+1}^K \, \, \sum\limits_{\ell \in r + 2 K \mathbb{Z}} \, f(\ell).$$ This reorganization gives:
\begin{align}
    F(z,y) &=\sum_{l = 0}^{L} h^l  \sum_{r=-K+1}^K \sum_{j = 0}^J  \sum\limits_{\ell \in r + 2 K \mathbb{Z}} a_{\ell,j,l}(\tau) \,e^{2\pi i \ell z} q^{\ell^2/(4K)}  \Bigr( y + \frac{\ell}{2K} \Bigr)^j.\label{eq:step2}
\end{align}
We therefore would be done, in case we can show 
\begin{equation}
    a_{r + 2K,j,l}(\tau) = a_{r,j,l}(\tau). \label{eq:period}
\end{equation}
Since then it would follow that for $\ell \in r + 2 K \mathbb{Z}$, $a_{\ell, j,l}(\tau) = a_{r,j,l}(\tau)$ allowing us to bring the $a_{r,j,l}(\tau)$ coefficient outside the $\ell$ summation, so that what remains naturally becomes the $\Th^{j}_{K,r}(z,y|\tau)$ function. So therefore we now prove \eqref{eq:period}. We go back to the expansion \eqref{eq:step1} and plug this into both sides of the second property \eqref{prop:gen}. The left hand side becomes
\begin{align}
    F(z+\tau, y+1) &= \sum_{l = 0}^{L} h^l  \sum_{l'=-\infty}^\infty \sum_{j = 0}^J  a_{l',j,l}(\tau) \,e^{2\pi i l'z} \, q^{l'} q^{l'^2/(4K)}  \Bigr( y + \frac{l'+2K}{2K} \Bigr)^j \label{eq:lhs}
\end{align}
where as the right hand side gives:
\begin{align}
    q^{-K} &e^{-4\pi i K z } F(z,y) = q^{-K} e^{-4\pi i K z}\sum_{l = 0}^{L} h^l  \sum_{l''=-\infty}^\infty \sum_{j = 0}^J  a_{l'',j,l}(\tau) \,e^{2\pi i l''z} q^{l''^2/(4K)}  \Bigr( y + \frac{l''}{2K} \Bigr)^j \nonumber \\
    &= q^{-K} e^{-4\pi i K z}\sum_{l = 0}^{L} h^l  \sum_{l'=-\infty}^\infty \sum_{j = 0}^J  a_{l'+2K,j,l}(\tau) \,e^{2\pi i (l'+2K)z} q^{(l'+2K)^2/(4K)}  \Bigr( y + \frac{l'+2K}{2K} \Bigr)^j \nonumber \\
    &= \sum_{l = 0}^{L} h^l  \sum_{l'=-\infty}^\infty \sum_{j = 0}^J  a_{l'+2K,j,l}(\tau) \,e^{2\pi i l'z} q^{l'^2/(4K)+ l'}  \Bigr( y + \frac{l'+2K}{2K} \Bigr)^j.\label{eq:rhs}
\end{align}
In the second equality we have simply translated the sum index by substituting $l'' = l' + 2K$. Now comparing the final equality above with \eqref{eq:lhs} we conclude $a_{l'+2K, j,l}(\tau) = a_{l',j,l}(\tau)$. By repeated applications we can always make $l'$ lie in between $-K+1$ and $K$, which we called $r$ earlier. This completes the proof. 
\end{proof}

\subsection{Theta orthogonality}
\label{app:lemorth}
We want to show here:
For residues $r,r'$ and integers $j,j'\ge0$,
\begin{equation}
  \int_{T_\tau}\dd^2z\;e^{-4\pi K\tau_2y^2}\;
  \Th^{(j)}_{K,r}(z,y)\;\overline{\Th^{(j')}_{K,r'}(z,y)}\;\Big|_{y=\Im z/\tau_2}
  = \delta_{r,r'}\;\tau_2\int_{-\infty}^{\infty}\dd u\;u^{\,j+j'}\,e^{-4\pi K\tau_2u^2} .
\end{equation}
\begin{proof}
Use the coordinates: $z=\sigma_1+\tau\sigma_2$, $y=\sigma_2$,
$\dd^2z=\tau_2\,\dd\sigma_1\dd\sigma_2$. Take the term $\ell$ of the first series and the
term $\ell'$ of the second. Their product contains
\begin{equation}
  e^{2\pi i\ell z}\,\overline{e^{2\pi i\ell' z}}
  = e^{2\pi i(\ell-\ell')\sigma_1}\;e^{2\pi i\sigma_2(\ell\tau-\ell'\bar\tau)} .
\end{equation}
The integral over $\sigma_1\in[0,1)$ gives $\delta_{\ell,\ell'}$, so in particular
$r=r'$. For $\ell=\ell'$ the remaining factors are
\begin{align}
  q^{\ell^2/(4K)}\,\bq^{\,\ell^2/(4K)}\;e^{-4\pi\ell\tau_2\sigma_2}\;e^{-4\pi K\tau_2\sigma_2^2}
  &= \exp\Bigl[-\frac{\pi\tau_2\ell^2}{K} - 4\pi\ell\tau_2\sigma_2 - 4\pi K\tau_2\sigma_2^2\Bigr]
  \nonumber\\
  &= \exp\Bigl[-4\pi K\tau_2\Bigl(\sigma_2+\frac{\ell}{2K}\Bigr)^2\Bigr] ,
\end{align}
together with the polynomial $(\sigma_2+\tfrac{\ell}{2K})^{j+j'}$. Note that the phases
$e^{\pm 2\pi i\tau_1\ell^2/(4K)}$ cancel. Now write $\ell=r+2Kt$ with $t\in\mathbb{Z}$; then
$u:=\sigma_2+\tfrac{\ell}{2K}=\sigma_2+\tfrac{r}{2K}+t$. As $\sigma_2$ runs over $[0,1)$ and $t$ runs
over $\mathbb{Z}$, $u$ runs over the real line exactly once. The sum over $t$ and the
integral over $\sigma_2$ therefore combine into one integral over $u\in\mathbb{R}$, which
gives \eqref{eq:orth}.
\end{proof}

\subsection{Modular properties of the one loop integrand}
\label{app:S-theta}

\subsubsection*{Modular invariance}
To find the modular transformation of the integrand $\mathcal{I}$ in \eqref{eq:Iblocks}, we investigate the  transformation properties of each building block. The objects $\mathcal{B}_{a,b}$ contain terms having the structure given in \eqref{B_ab-structure}. An arbitrary $SL(2,\mathbb{Z})$ transformation can be defined by $\Gamma=\left(\begin{smallmatrix}A&B\\C&D\end{smallmatrix}\right)$ with $AD-BC=1$, $C>0$. This  leads to $z \to z'=\frac{z}{C \tau +D}$ and $\tau \to \tau' = \frac{A \tau +B}{C \tau +D}$. The completed bosonic contractions $G_n,R_n$ have weight $n$ under this change, while $S_{s,n},T_{s,n}$ have weight $n+1$ with the permutation $\sigma$ exchanging $s=2$ and $s=4$. But the spin summed product of Szeg\H{o} kernels transforms nicely with a fixed modular weight. Using these we can determine the weight of the block $\mathcal{B}_{a,b}$ under arbitrary modular transformation.
\begin{align}
    &G'_n(z',\tau') = (C \tau + D)^n G_n(z,\tau); \nonumber \\
    &\sum_s w'_s \prod_{i=1}^{2k}S'_{s,n_i} = (C \tau + D)^{-4+\sum_i(n_i+1)} \sum_s w_s \prod_{i=1}^{2k}S_{s,n_i}; \nonumber \\
    &\mathcal{B}'_{a,b} = (C \tau + D)^{W-4} \mathcal{B}_{a,b}. 
\end{align}
Using modular transformation of $e^{-G}$ one can find the modular transformation of the full integrand.
\begin{align}
    \mathcal{B}_{a,b} \overline{\mathcal{B}_{a,b}}|E|^{4K} e^{-4 \pi K \tau_2 y^2} \to |C \tau + D|^{2 W -8 - 4 K} \mathcal{B}_{a,b} \overline{\mathcal{B}_{a,b}}|E|^{4K} e^{-4 \pi K \tau_2 y^2}.
    \label{B-transformation}
\end{align}
Modular transformations of $h$ and the integrand measure is given by,
\begin{align}
    h \to |C \tau + D|^{2} h, \qquad \frac{\text{d}^2 \tau ~ \text{d}^2 z}{\tau_2^5} \to |C \tau + D|^{4} \frac{\text{d}^2 \tau ~ \text{d}^2 z}{\tau_2^5}.
\end{align}
Using the explicit value of $W$ in \eqref{block-product} we see the full integrand of \eqref{eq:Adef} (along with the integration measure) is invariant under modular transformation.

\subsubsection*{Modular transformation of $\Theta$ functions}

The basis of the $SL(2,\mathbb{Z})$ transformations are the $S$ transformation and $T$ transformation. 
\paragraph{$S$ transformation} is defined as $\tau \rightarrow \tau'=-1/\tau$ and $z \rightarrow z'=z/\tau$. Under $S$ transformation the functions $\Theta^{(0)}_{K,r}$ transforms as,
\begin{align}
    \Theta^{(0)}_{K,r}(z',y'|\tau') &=  \sqrt{\frac{-i\tau}{2K}} e^{2\pi iKz^2/\tau}\times
 \sum_{\rho\,(\mathrm{mod}\,2K)}e^{-\pi ir\rho/K}\Theta^{(0)}_{K,\rho}(z,y|\tau).
\end{align}
\paragraph{$T$ transformation} takes $\tau$ to $\tau'= \tau+1$, keeping the coordinate $z$ unchanged. Under $T$ transformation the $\Theta$ functions transform as 
\begin{align}
    \Theta^{(0)}_{K,r}(z'|\tau') &=  e^{\pi ir^2/2K}\Theta^{(0)}_{K,r}(z|\tau).
\end{align}
The modular transformation $\Gamma$ can be written as a sequence of $S$ and $T$ transformations. Using these one can determine the transformation properties of the functions $\Theta^{(0)}_{K,r}$ under generic modular transformations. To get the modular transformations of $\Theta^{(i)}_{K,r}$ we consider a generating functional of the form
\begin{align}
    \mathcal{G}_r(t,z|\tau) =\sum_i \tfrac{t^i}{i!}\Theta^{(i)}_{K,r}(z,y|\tau) = e^{ty}\Theta^{(0)}_{K,r}(z+t/(4\pi iK)|\tau)
\end{align}
Using the modular transformation of the generating functional $\mathcal{G}$ we obtain the modular transformation of the $\Theta^{(i)}_{K,r}$ functions also.
\begin{align}
    &\Theta^{(0)}_{K,r}(z'|\tau') = e^{\frac{i 2 \pi K C z^2}{(C \tau +D)}} \sum_{r' = -K+1}^K \mathcal{M}^{(0)}_{r,r'}(\Gamma,\tau) \Theta^{(0)}_{K,r'}(z|\tau); \nonumber \\
    &\Theta^{(i)}_{K,r}(z',y'|\tau') = e^{\frac{i 2 \pi K C z^2}{(C \tau +D)}} \sum_{r' = -K+1}^{K} \sum_{i'=0}^{i}  \mathcal{M}^{i',r'}_{i,r} \Theta^{(i')}_{K,r'}(z,y|\tau); \nonumber \\
    &\text{where,} \qquad \mathcal{M}^{i',r'}_{i,r} = \begin{cases}
        \mathcal{M}^{(0)}_{r,r'}(\Gamma,\tau) \tfrac{(C\tau + D)^{(i+i')/2} i!}{\left( \tfrac{i-i'}{2} \right)!i'!} \left( \tfrac{-i C}{8 \pi K} \right)^{\tfrac{i-i'}{2}} \qquad &\text{for } i -i'\equiv 0(\text{mod}2) \\
        0 \qquad &\text{else} 
    \end{cases}.
\end{align}
Here, the matrix $\mathcal{M}^{(0)}_{r,r'}$ is independent of the coordinate $z$. It only depends on the parameters of the modular transformation. We know the modular properties of the integrand as a whole, that is 
\begin{align}
    \Phi_{a,b}(z',y') \bar{\Phi}_{a,b} (\bar{z}',y') e^{- 4 \pi K \tau'_2 y'^2} = \Phi_{a,b}(z,y) \bar{\Phi}_{a,b} (\bar{z},y) e^{- 4 \pi K \tau_2 y^2} |C \tau + D|^{-4-2|a|-2|b|}.
    \label{integrand:transformation}
\end{align}
The exponential factor that comes from the transformation of the theta functions, gives a modular invariant combination together with the exponential factor coming from the antiholomorphic part and the exponential measure of the integration.
\begin{align}
    e^{\frac{i 2 \pi K C z^2}{(C \tau +D)}} e^{\frac{-i 2 \pi K C \bar{z}^2}{(C \bar{\tau} +D)}} e^{- 4 \pi K \tau'_2 y'^2}= e^{- 4 \pi K \tau_2 y^2}.
    \label{exponential:transformation}
\end{align}
The blocks $\Phi_{a,b}$ can be expanded in terms of the theta functions. 
\begin{equation}
 \Phi_{a,b}(z,y|\tau)
 =
 \sum_{r,i,l}
 h^l a^{(a,b)}_{r,i,l}(\tau)
 \Theta^{(i)}_{K,r}(z,y|\tau).
 \label{eq:S-block-expansion}
\end{equation}
If we consider that the coefficients $a$ transform with a matrix $\rho$, then using \eqref{integrand:transformation} and \eqref{exponential:transformation} we can write
\begin{align}
    &\left( \sum_{i,i'} \sum_{r,r'}\left( \rho^{i',r'}_{i,r} a^{a,b}_{i',r'}(\tau)\right) \left( \mathcal{M}^{i',r'}_{i,r} \Theta^{(i')}_{K,r'}(z,y|\tau) \right) \right) 
    \left( \sum_{j,j'} \sum_{s,s'}\left( \tilde{\rho}^{j',s'}_{j,s} \tilde{a}^{b,a}_{j',s'}(\bar{\tau})\right) \left( \tilde{\mathcal{M}}^{j',s'}_{j,s} \Theta^{(j')}_{K,s'}(\bar{z},y|\bar{\tau}) \right) \right) \nonumber \\
    &= \left( \sum_{ir}  a^{a,b}_{i,r}(\tau) \Theta^{(i)}_{K,r}(z,y|\tau) \right) \left( \sum_{j,s} \tilde{a}^{b,a}_{j,s}(\bar{\tau}) \Theta^{(j)}_{K,s}(\bar{z},y|\bar{\tau}) \right) |C \tau + D|^{-4-2|a|-2|b|}.
\end{align}
Hence, the transformation matrix $\rho$ and $\tilde{\rho}$ are determined in terms of the inverse of the transformation matrices $\mathcal{M}$:
\begin{align}
    &\rho^{i',r'}_{i,r} (\Gamma, \tau) = \left(\mathcal{M}^{-1}\right)^{i',r'}_{i,r} \left(C \tau + D\right)^{-2-|a|-|b|}, \nonumber \\
    &\tilde{\rho}^{i',r'}_{i,r} (\Gamma, \bar{\tau}) = \left( \tilde{\mathcal{M}}^{-1}\right)^{j',s'}_{j,s} \left(C \bar{\tau} + D\right)^{-2-|a|-|b|}.
    \label{eq:rho rhotilde}
\end{align}

\section{Numerical details and error analysis}
\label{app:numdet}

There are two stages at which numerical errors can enter. We first
extract the coefficients of the theta expansion and truncate their
$q$-series. We then integrate over the modulus. The examples below
allow us to check the two stages separately, since at $K=1$ we also
have the chiral factor and the insertion integral in closed form.

\subsection{Coefficient extraction and series truncation}
\label{app:coefficientsector}
\label{app:extraction}

Let $P(z,y,h;\tau)$ denote a chiral factor, with the partition and
removal labels suppressed. We sample $P(x_k,y,h;\tau)$ at
$x_k=(k+\tfrac12)/N_z$, $0\le k<N_z$, treating $y,h$ as independent.
The half-step avoids the poles of the individual contraction functions
at $z=0$. A discrete Fourier transform in $x$ gives
\begin{equation}
 \widetilde P_r(y,h;\tau)=\frac{1}{N_z}
 \sum_{k=0}^{N_z-1}e^{-2\pi irx_k}P(x_k,y,h;\tau).
\end{equation}
We expand $\widetilde P_r(u-r/(2K),h;\tau)
=\sum_{j,l}\widetilde b_{r,j,l}(\tau)u^jh^l$ as in
\eqref{eq:coeffformula}. One can now expand,
\begin{equation}
 \widetilde b_{r,j,l}(\tau)
 =\sum_{n\in\mathbb Z}c_{r,j,l,n} q^n.
 \label{eq:Cbseries}
\end{equation}
At a fixed extraction height $\tau_h$, the
$q$-series coefficients are then approximated by
\begin{equation}
 \widetilde c_{r,j,l,n}
 =\frac{e^{2\pi n\tau_h}}{N_q}
 \sum_{v=0}^{N_q-1}
 e^{-2\pi inv/N_q}\widetilde b_{r,j,l}(v/N_q+i\tau_h).
 \label{eq:Cextract}
\end{equation}
The exponential prefactor removes the damping of the $n$th mode on
the extraction circle. We use the coefficients for $0\le n\le N<N_q$ in \eqref{eq:Cbseries}, and hence
in \eqref{eq:aexpansion}. Refining $N_z$ and $N_q$ tests the two
sampling errors. Varying $N$ at fixed grids tests the discarded tail.

\subsubsection{Coefficient checks}
\label{app:testfunctions}

At $K=1$ the physical chiral factor is $P_0=E^2$. To test a
$y$-dependent block we also use $P_1=E^2G_1$. Equation~\eqref{eq:Drule}
and $P_1=-DP_0/2$ imply that the $P_1$ expansion has only $j=1$,
with coefficients $-2\pi i$ times the $j=0$ coefficients of $P_0$.
The theta-constant expression in Appendix~\ref{app:K1-zintegral}
therefore gives reference coefficients for both tests.

We compare nonzero coefficients using the maximum relative error in
each transform. At $\tau=0.23+0.45i$, increasing $N_z$ from 4 to 12
and then to 24 reduces the error from $3.50\times10^{-3}$ to
$1.45\times10^{-37}$ and then to $2.62\times10^{-81}$ in 80-digit
arithmetic. At $\tau_h=0.45$, increasing $N_q$ from 12 to 32 and 64
reduces the $q$-coefficient error from $1.41\times10^{-9}$ to
$6.30\times10^{-29}$ and $2.16\times10^{-62}$.

\begin{figure}[htbp]
 \centering
 \includegraphics[width=\textwidth]{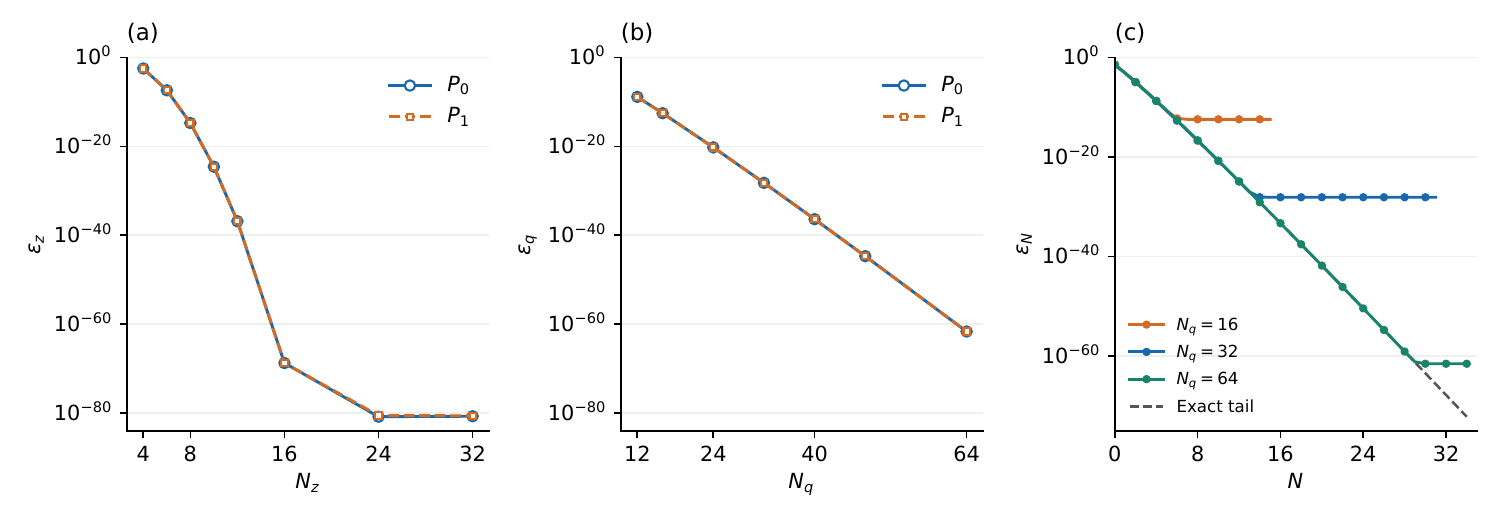}
 \caption{Maximum relative errors in the $z$-Fourier coefficients (a),
 $q$-series coefficients (b), and $P_1$ reconstruction (c).
 Panel (c) holds $N_z=16$ and shows $N_q=16,32,64$ with $N<N_q$.
 The dashed curve uses exact coefficients. The working precision is
 80 digits and the $q$-extraction height is $\tau_h=0.45$.}
 \label{fig:Cdependences}
\end{figure}

For fixed extraction grids, increasing $N$ suppresses the omitted
tail until the error approaches the accuracy of the retained
coefficients. At $N_z=16$, $N_q=32$ and $\tau_h=0.45$, the maximum
relative reconstruction errors sampled where the reference is
nonzero are as follows.
\begin{center}
\small
\begin{tabular}{cccc}
 \toprule
 $N$ & $P_0$ total & $P_1$ total & $P_1$ exact-coefficient tail\\
 \midrule
 $0$  & $2.95\times10^{-2}$  & $3.73\times10^{-2}$  & $3.73\times10^{-2}$\\
 $4$  & $1.39\times10^{-9}$  & $1.93\times10^{-9}$  & $1.93\times10^{-9}$\\
 $8$  & $1.52\times10^{-17}$ & $2.18\times10^{-17}$ & $2.18\times10^{-17}$\\
 $12$ & $8.77\times10^{-26}$ & $1.28\times10^{-25}$ & $1.28\times10^{-25}$\\
 $16$ & $6.46\times10^{-29}$ & $8.40\times10^{-29}$ & $5.11\times10^{-34}$\\
 $24$ & $6.46\times10^{-29}$ & $8.40\times10^{-29}$ & $3.89\times10^{-51}$\\
 \bottomrule
\end{tabular}
\end{center}
Figure~\ref{fig:Cdependences} shows the two grid refinements and how
the reconstruction error approaches this limit at different $N_q$.

\subsection{Modular integration checks}
\label{app:validation_scope}

We use the level-one integrand in \eqref{eq:A1} to test the
modular integration independently of coefficient extraction. We
evaluate the theta constants directly on the compact domain and
use exact series coefficients in the cusp. At $T=1$, extending
the cusp series from $N=8$ to $N=64$ changes $A$ by
$2.52\times10^{-44}$. In the following tests $N=64$.

\paragraph{Compact quadrature.}
Write $Q_{L_1,L_2}$ for the compact integral with $L_1$ and $L_2$
Gauss--Legendre nodes in $\tau_1$ and $\tau_2$. We monitor
\begin{equation}
 \Delta_1=|Q_{2L_1,L_2}-Q_{L_1,L_2}|,\qquad
 \Delta_2=|Q_{L_1,2L_2}-Q_{L_1,L_2}|.
\end{equation}
Refinement in $\tau_1$ controls initial convergence.
At $T=1$, $Q_{64,64}$ differs from an independent 50-digit
reference by $2.16\times10^{-18}$.

\paragraph{Cusp kernels.}
For $\kappa=4\pi m$, the recurrence is
\begin{equation}
 J_s(m;T)=\frac{T^{1-s}e^{-\kappa T}-\kappa J_{s-1}(m;T)}{s-1}.
 \label{eq:Ckernelrecurrence}
\end{equation}
For $m<0$ its starting value with the $-i0$ prescription is
\begin{equation}
 J_{1/2}(m;T)=
 \sqrt{\frac{\pi}{-\kappa}}\,
 \left[i-\operatorname{erfi}\bigl(\sqrt{-\kappa T}\bigr)\right].
\end{equation}
For $s=9/2$, $m=-1/4$ and $T=1,1.3,1.5,1.7,2$, the maximum
difference $\delta J_s$ from
$T^{1-s}E_s(4\pi mT-i10^{-45})$, evaluated at 50-digit precision,
is $1.80\times10^{-44}$.

\begin{figure}[htbp]
 \centering
 \includegraphics[width=\textwidth]{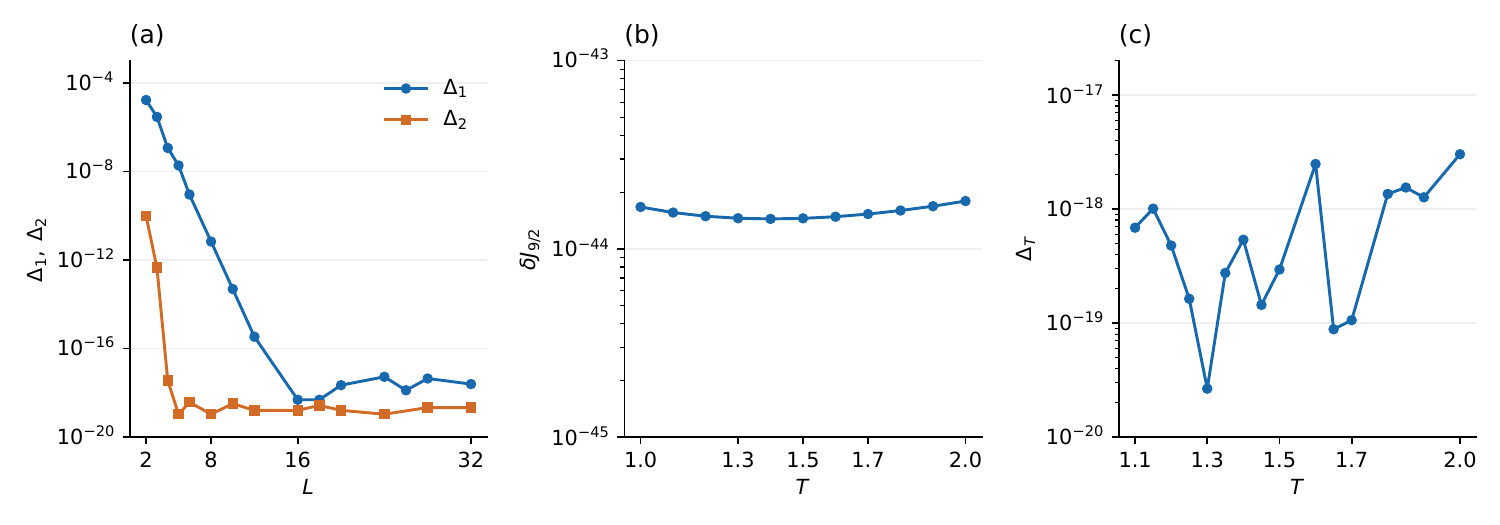}
 \caption{Modular-integration checks for $K=1$.
 (a) $\Delta_1$ and $\Delta_2$ versus the initial node count $L$, at $T=1$.
 (b) $\delta J_{9/2}$ versus $T$ for $m=-1/4$, at 50-digit precision.
 (c) $\Delta_T$ versus $T$, with $L_1=L_2=64$ and $N=64$.}
 \label{fig:Cmodularchecks}
\end{figure}

\paragraph{Splitting height.}
Let $\Delta_T$ be the difference in the total amplitude between
height $T$ and height 1. With $L_1=L_2=64$ and 50-digit cusp
kernels, the largest absolute difference for
$T=1.1,1.2,1.3,1.5,1.7,2$ is $3.02\times10^{-18}$.
Figure~\ref{fig:Cmodularchecks} shows the three integration tests.

\section{Rademacher evaluation of the modular integral}
\label{app:rademacher}

With the modular transformation results derived in appendix \ref{app:S-theta}, we are now ready to use the Rademacher technology of \cite{BacciantiEtAl2025}.

	For fixed $(a,b),$ we denote the
	external-state and combinatorial prefactor by $W_{ab}$. The original integral is then,

    \[
     \mathcal A_{ab}=W_{ab}\int_{\mathcal F}\frac{d^2\tau}{\tau_2^5}\,h^{|a|+|b|}
     \int_{T_\tau}d^2z\,e^{-4\pi K\tau_2 y^2}
           \Phi_{a,b}(z,y|\tau)\,\overline{\Phi_{b,a}(z,y|\tau)} .
    \]
	Now use the expansion from \eqref{eq:S-block-expansion}, 
    \begin{align}
        \Phi_{a,b}=\sum_{r,j,l}
    h^l a^{(a,b)}_{r,j,l}(\tau)
    \Theta^{(j)}_{K,r}(z,y|\tau), \qquad 
        \overline{\Phi_{b,a}(z,y\mid\tau)}
    =
    \sum_{r',j',l'}
    h^{\,l'}\,
    \overline{a^{(b,a)}_{r',j',l'}(\tau)}\,
    \overline{\Theta^{(j')}_{K,r'}(z,y\mid\tau)} .
    \end{align}
 Using the orthogonality of the $\Theta$ functions mentioned in \eqref{eq:orth}, we obtain,

    \begin{align}
        \mathcal A_{ab}
={}& \sum_{\substack{r,j,l,j',l'\\j+j'\ {\rm even}}}
\widetilde{W}^{(a,b)}_{j,j',l,l'}
\int_{\mathcal{F}} d^2\tau \, \tau_2^{\,-s^{(a,b)}_{l,l',j,j'}}
\,a^{(a,b)}_{r,j,l}(\tau)\,
\overline{a^{(b,a)}_{r,j',l'}(\tau)}, 
    \end{align}
    where,
    \begin{equation}
    \widetilde{W}^{(a,b)}_{j,j',l,l'}=\frac{W_{ab}\pi^{|a|+|b|+l+l'}(j+j'-1)!!}
     {2\sqrt{K}\,(8\pi K)^{(j+j')/2}}, \qquad 
     s^{(a,b)}_{l,l',j,j'}
=
\frac{9}{2} + |a| + |b| + l + l' + \frac{j+j'}{2}.
    \end{equation}

    Following \cite{BacciantiEtAl2025}, for the independent right-moving parameter $\tilde\tau$ we now define
    $\tilde a^{(b,a)}_{r,j',l'}(\tilde\tau)
    =\overline{a^{(b,a)}_{r,j',l'}(-\overline{\tilde\tau})}$.
    On the physical slice $\tilde\tau=-\bar\tau$ one has
    $(\tau+\tilde\tau)/(2i)=\tau_2$ and $d\tau\wedge d\tilde\tau/(2i)=d^2\tau$.
    The two-form corresponding to Eq.~(3.1) of \cite{BacciantiEtAl2025} is then,
    \begin{equation}
     \omega^{(a,b)}_{r,j,j',l,l'}=\frac{d\tau\wedge d\tilde\tau}{2i}
     \left(\frac{\tau+\tilde\tau}{2i}\right)^{\,-s^{(a,b)}_{l,l',j,j'}}
     a^{(a,b)}_{r,j,l}(\tau)\,\tilde a^{(b,a)}_{r,j',l'}(\tilde\tau),
     \label{radf:31}
    \end{equation}
    and the full integral is,
    \begin{equation}
        \mathcal A_{ab}
    =\sum_{\substack{r,j,l,j',l'\\j+j'\ {\rm even}}}
    \widetilde{W}^{(a,b)}_{j,j',l,l'} \int\omega^{(a,b)}_{r,j,j',l,l'}\left. \right|_{\tilde\tau=-\bar\tau}.
    \end{equation}
	
    \noindent Under a modular transformation with $\Gamma=\left(\begin{smallmatrix}A&B\\C&D\end{smallmatrix}\right)$, the $\tilde{a}$ factor transforms as, 
    \begin{equation}
     \tilde a^{(b,a)}_{r,j',l'}(\Gamma\tilde\tau)=
     \sum_{r'',j'',l''}
     \frac{(\tilde\rho^\Gamma)_{j',r,l'}^{j'',r'',l''}}
          {[-i(C\tilde\tau+D)]^{\,s-\nu-2}}\,
     \tilde a^{(b,a)}_{r'',j'',l''}(\tilde\tau).
     \label{radf:32}
    \end{equation}
    In the above expression, $s\equiv s^{(a,b)}_{l,l',j,j'}$ and $\nu=(j-j'')/2+l-l''$; we have suppressed the indices for brevity. The $\tilde\rho^\Gamma$ can be expressed in terms of the $\tilde\rho$ factors in \eqref{eq:rho rhotilde},
    \begin{equation}
     (\tilde\rho^\Gamma)_{j',r,l'}^{j'',r',l''}
     =(-iJ)^{-s+\nu+2}
     \tilde\rho_{j',r}^{\,j'',r'}(\Gamma^\#,-\tilde\tau)
     \binom{l''}{l'}(2\pi i C)^{l''-l'} J^{-l'-l''},
    \end{equation}
    with $J=C\tilde\tau+D$ and
    $\Gamma^\#=\left(\begin{smallmatrix}A&-B\\-C&D\end{smallmatrix}\right)$.

    Let $\mathtt{C}_{A/C}$ be the clockwise Ford circle at the reduced cusp $A/C$,
with $C>0$ and $AD-BC=1$. Write $\to$ for a rightward horizontal contour
and $s(A,C)$ for the Dedekind sum. By  \cite{BacciantiEtAl2025},
the contribution of this cusp to the principal-value integral is,
\begin{equation}
 \mathcal I^{(a,b)}_{A/C}
 =\frac16\sum_{\substack{r,j,l,j',l'\\j+j'\ {\rm even}}}
 \widetilde W^{(a,b)}_{j,j',l,l'}
 \int_{\tau\in\to}\int_{\tilde\tau\in \mathtt{C}_{A/C}}
 \left(\tau-\tilde\tau+\frac{2A}{C}-12s(A,C)\right)
 \omega^{(a,b)}_{r,j,j',l,l'} .
 \label{radf:33}
\end{equation}
Then the transformation \eqref{radf:32} leads to,
\begin{equation}
 \mathcal I^{(a,b)}_{A/C}
 =\sum_{\substack{r,j,l,j',l'\\j+j'\ {\rm even}}}
   \sum_{r'',j'',l''}
 \widetilde W^{(a,b)}_{j,j',l,l'}
 (\tilde\rho^\Gamma)_{j',r,l'}^{j'',r'',l''}\,
 \mathcal I_{A/C}^{\,rjl;\,rj'l';\,r''j''l''},
 \label{radf:split}
\end{equation}
with,
\begin{equation}
 \begin{aligned}
 \mathcal I_{A/C}^{\,rjl;\,rj'l';\,r''j''l''}
 ={}&\frac{2^{s-1}}{6i}\int_\to d\tau\int_\to d\tilde\tau\,
 \frac{[-i(C\tilde\tau+D)]^\nu\,
 a^{(a,b)}_{r,j,l}(\tau)\,\tilde a^{(b,a)}_{r'',j'',l''}(\tilde\tau)}
 {[-(A\tilde\tau+B+\tau(C\tilde\tau+D))]^s}\\
 &{}\times
 \left(\tau-\frac{A\tilde\tau+B}{C\tilde\tau+D}+\frac{2A}{C}-12s(A,C)\right).
 \end{aligned}
 \label{radf:34}
\end{equation}

Use the Fourier coefficients $c^{(a,b)}_{r,j,l,n}$ of \eqref{eq:aexpansion}. For the coefficient
pair in \eqref{radf:34}, the two Fourier exponents and the Bessel argument are
\[
 m=n-\frac{r^2}{4K},\qquad
 \bar m=n'-\frac{(r'')^2}{4K},\qquad
 X=\frac{4\pi\sqrt{|m\bar m|}}{C}.
\]
The horizontal integrals select $m,\bar m<0$. Using \cite{BacciantiEtAl2025}, one can directly evaluate the fixed-label result:
\begin{equation}
\begin{aligned}
 \mathcal I_{A/C}^{\,rjl;\,rj'l';\,r''j''l''}
 ={}&\frac{2^s\pi^2(2\pi)^{s-1}}{3C^2\Gamma(s)}
 \sum_{\substack{n,n'\ge0\\m,\bar m<0}}
 c^{(a,b)}_{r,j,l,n}\,
 \overline{c^{(b,a)}_{r'',j'',l'',n'}}\,
 e^{-2\pi i(Am+D\bar m)/C}\\
 &{}\times |m|^{(s+\nu-1)/2}|\bar m|^{(s-\nu-1)/2}
 \bigg[\left(12iC\,s(A,C)+\frac{C\nu}{2\pi|m|}\right)J_{s-\nu-1}(X)\\
 &\hspace{31mm}{}+
 \sqrt{\frac{|\bar m|}{|m|}}\,
 \big(J_{s-\nu-2}(X)-J_{s-\nu}(X)\big)\bigg].
\end{aligned}
\label{radf:312}
\end{equation}
Substituting the fixed-label result \eqref{radf:312} into \eqref{radf:split}
and summing over reduced cusps gives the principal value at fixed $(a,b)$, one obtains the full integral answer:
\begin{equation}
\displaystyle
 \operatorname{PV}\mathcal A_{ab}
 =\sum_{C=1}^{\infty}\ \sum_{\substack{0\le A<C\\(A,C)=1}}
  \sum_{\substack{r,j,l,j',l'\\j+j'\ {\rm even}}}
  \sum_{r'',j'',l''}
  \widetilde W^{(a,b)}_{j,j',l,l'}\,
  (\tilde\rho^\Gamma)_{j',r,l'}^{j'',r'',l''}\,
  \mathcal I_{A/C}^{\,rjl;\,rj'l';\,r''j''l''}.
\label{radf:amplitude}
\end{equation}

Table~\ref{tab:rademacher-checks} compares selected reduced amplitudes
with the compact-region quadrature plus continued cusp integral of
Sec.~\ref{sec:tauint}. For $K=1$, the normalization of
Ref.~\cite[Sec.~3.3]{BacciantiEtAl2025} gives $\pi^2 A_{(1)}$.
Both methods give $\Im A_{(1)}=1/210$, and the $K=9$ absorptive
entries agree to the displayed precision. The real parts differ by
$4.49\times10^{-10}$ and $9.62\times10^{-17}$ for the two $K=9$
entries, respectively, before rounding the table. Here $c_{\max}$ is
the largest denominator retained in the Rademacher sum.

\begin{table}[htbp]
 \centering
 \small
 \begin{tabular}{c l r r r r}
 \toprule
 $K$ & Entry & $c_{\max}$ & $\Re A$ (direct) &
 $\Re A$ (Rademacher) & $\Im A$ \\
 \midrule
 1 & $A_{(1)}$ & 1000 & $0.00223337755$ & $0.00223337770$ & $1/210$ \\
 9 & $|1^9,1^9\rangle$ diagonal & 2048 &
 $0.00722077763$ & $0.00722077808$ & $0.00802324997$ \\
 9 & Exchange off-diagonal & 2048 &
 $0.00082053868644$ & $0.00082053868644$ & $0.00174355813$ \\
 \bottomrule
 \end{tabular}
 \caption{Rademacher checks on selected matrix elements of the reduced
 amplitude $A$. The exchange entry connects $|1^9,(2,1^7)\rangle$
 and $|(2,1^7),1^9\rangle$.}
 \label{tab:rademacher-checks}
\end{table}
\FloatBarrier

\section{Spin multiplets and the DDF matrix}
\label{subsec:ddf_spin_degeneracy}

If a level contains $m_R$ copies of the little-group representation
$V_R$, a rotation-invariant self-energy acts on
$V_R\otimes\mathbb C^{m_R}$ as $\mathbb I_{V_R}\otimes\Sigma_R$.
The analogous statement for supersymmetry applies to long multiplets.
A DDF basis makes the oscillator structure explicit. Its matrix
elements in a fixed-polarization subspace are projections of the
full self-energy onto that subspace.

It is instructive to see how the same mass splitting looks in the
two bases. The first few levels of the open bosonic string give a
simple example. At $N=2$, the transverse vector
and symmetric tensor combine into the single $\mathbf{324}$ of
$SO(25)$, so the self-energy at that level is proportional to the
identity. At $N=3$, the states form the symmetric traceless
$\mathbf{2900}$ and the antisymmetric $\mathbf{300}$. Each
contains an $SO(24)$ vector. In the orthonormal DDF basis
\begin{equation}
 |u_i\rangle=\frac{1}{\sqrt3}A_{-3}^i|p\rangle,\qquad
 |v_i\rangle=\frac{1}{\sqrt{52}}(A_{-1}\cdot A_{-1})A_{-1}^i|p\rangle,
\end{equation}
the two vector directions belonging to the two $SO(25)$ multiplets are
\begin{equation}
 |S_i\rangle=\frac{\sqrt{13}|u_i\rangle+\sqrt3|v_i\rangle}{4},
 \qquad
 |A_i\rangle=\frac{\sqrt3|u_i\rangle-\sqrt{13}|v_i\rangle}{4}.
\end{equation}
If the two multiplets have shifts $\lambda_S,\lambda_A$, the vector
block in the DDF basis is
\begin{equation}
 \frac1{16}
 \begin{pmatrix}
 13\lambda_S+3\lambda_A&\sqrt{39}(\lambda_S-\lambda_A)\\
 \sqrt{39}(\lambda_S-\lambda_A)&3\lambda_S+13\lambda_A
 \end{pmatrix}.
\end{equation}
This off-diagonal matrix element encodes the splitting of the two
spin multiplets.

\subsection{Long multiplets in Type II}
\label{subsec:typeII_degeneracy}

In flat ten dimensions the chiral first massive level is a
256-state multiplet
$\mathcal V=(\mathbf{44}\oplus\mathbf{84})_{\mathrm B}
\oplus\mathbf{128}_{\mathrm F}$. The NS part has
$8+8^2+\binom83=128$ states from the oscillator structures
$B_{-3/2}$, $A_{-1}B_{-1/2}$ and $B_{-1/2}^3$.
The closed-string level is one long multiplet
$\mathcal U=\mathcal V_L\otimes\mathcal V_R$, of dimension $65536$,
with a common complex pole shift. At $K=2$ and $K=3$, the remaining
little-group representations after factoring out $\mathcal U$ are
\cite{BianchiMoralesSamtleben2003,ForcellaHananyTroost2010,PasquaZumino2004}
\begin{align}
 \mathcal R_2&=\mathbf 9\otimes\mathbf 9,\\
 \mathcal R_3&=(\mathbf{44}_{\mathrm B}\oplus\mathbf{16}_{\mathrm F})
              \otimes(\mathbf{44}_{\mathrm B}\oplus\mathbf{16}_{\mathrm F}).
\end{align}
For example, at $K=2$ the chiral NS space has dimension $1152$ and
transforms as $\mathbf 9\otimes(\mathbf{44}\oplus\mathbf{84})$.
Supersymmetry makes $\Sigma$ the identity on $\mathcal U$ times an
operator on $\mathcal R_K$. The labels in $\mathcal R_K$ are
superspins of long multiplets.

At $K=2$ the three superspins are
$\mathbf 9\otimes\mathbf 9=\mathbf1\oplus\mathbf{36}\oplus\mathbf{44}$.
At $K=3$ write $H=\mathbf{44}$ and $S=\mathbf{16}$. The chiral NS/R
selection rule can be expressed on the residual space by multiplying
each NS/R parity by the chirality of its normalized supercharge
Clifford algebra. Each factor anticommutes with the chiral supercharges,
so the product commutes with them. It distinguishes $H$ from $S$.
The two-point function respects this parity in each chirality, and
left--right exchange relates $H\otimes S$ to $S\otimes H$. The
remaining product decompositions are
\begin{align}
 H\otimes H &=\mathbf1\oplus\mathbf{36}\oplus\mathbf{44}
              \oplus\mathbf{495}\oplus\mathbf{910}\oplus\mathbf{450},\nonumber\\
 S\otimes S &=\mathbf1\oplus\mathbf9\oplus\mathbf{36}
              \oplus\mathbf{84}\oplus\mathbf{126},\nonumber\\
 H\otimes S &=\mathbf{128}\oplus\mathbf{576}.
 \label{eq:residual-K3}
\end{align}
Thus five $S\otimes S$ and two $H\otimes S$ pole shifts are
independent of the six in $H\otimes H$.

\subsection{Reconstruction of the low-level shifts}

Let $E_K$ embed the chiral DDF states into the residual representation
with the common $\mathcal U$ state suppressed, and let $P_R$ be the
projector onto residual irrep $R$. If $\mu_R$ is its complex
mass-squared shift, the restricted matrix is
\begin{equation}
 \Sigma_{\mathrm{DDF}}=
 \sum_R\mu_R\,(E_K\otimes E_K)^\dagger P_R(E_K\otimes E_K).
 \label{eq:low-compression}
\end{equation}
On each multiplicity-free product, the $P_R$ may be formed from the
quadratic $\mathrm{Spin}(9)$ Casimir $\mathcal C$ as
$P_R=\prod_{R'\ne R}(\mathcal C-c_{R'})/(c_R-c_{R'})$, with $c_R$ the
Casimir eigenvalue. Because $E_K$ spans only part of the product
representation, the pole shifts are recovered by solving the linear
equations for $\mu_R$.

In light-cone variables the chiral bosonic DDF modes are the
transverse oscillators. Write $e_9$ for the extra spatial axis
of the massive little group. At $K=2$ the normalized states for
$a=(1,1)$ and $b=(2)$ embed as $\lambda$ and $e_9$ in $\mathbf9$.
At $K=3$ the states for $a=(1,1,1)$, $b=(2,1)$ and $c=(3)$ embed
in $H$ as the symmetric traceless tensors
\begin{equation}
 \lambda_I\lambda_J,\qquad
 \frac{\lambda_I(e_9)_J+(e_9)_I\lambda_J}{\sqrt2},\qquad
 \frac{9(e_9)_I(e_9)_J-\delta_{IJ}}{\sqrt{72}}.
 \label{eq:residual-embedding}
\end{equation}
These tensors are the transverse highest weights and the single-mode
singlet in the branching of $H$ under $SO(8)$. The massless seed
polarization is common to the three chiral states.

For partitions $p\ne q$, write
$F_{pq}^{\pm}=\langle p,q|\Sigma|p,q\rangle
\pm\langle p,q|\Sigma|q,p\rangle$ and
$F_{pp}=\langle p,p|\Sigma|p,p\rangle$. At $K=2$, the singlet,
antisymmetric and symmetric traceless projectors give
\begin{equation}
 \mu_1=9F_{bb}-8F_{aa},\qquad \mu_{36}=F_{ab}^-,\qquad \mu_{44}=F_{aa}.
 \label{eq:K2-poles}
\end{equation}
The exchange-even entry satisfies $F_{ab}^+=F_{aa}$.
At $K=3$, inserting \eqref{eq:residual-embedding} into the
$H\otimes H$ projectors gives
\begin{align}
 \mu_{450}&=F_{aa},&
 \mu_{910}&=F_{ab}^{-},\nonumber\\
 \mu_{44}&=\tfrac17(52F_{bc}^{+}-45\mu_{450}),&
 \mu_{36}&=\tfrac19(44F_{bc}^{-}-35\mu_{910}),\nonumber\\
 \mu_{495}&=\tfrac1{26}(91F_{bb}-9\mu_{44}-56\mu_{450}),&
 \mu_1&=44F_{cc}-\tfrac{154}{13}\mu_{44}-\tfrac{405}{13}\mu_{450}.
 \label{eq:K3-poles}
\end{align}
The other entries of the restricted matrices can be used to check the
reconstruction. Table~\ref{tab:low-poles} lists the resulting shifts
and widths. The seven other shifts at $K=3$ require states with a
residual spinor in at least one chirality.

\section{Independence of the PCO positions}
\label{app:spurious-pco}

At loop level, spurious poles can arise when a picture-changing operator
(PCO) meets another PCO or a vertex, or when a superghost theta function
vanishes on a smooth surface. Consider the torus two-point amplitude
\begin{equation}
\mathcal A=\int_{\mathcal F}d^2\tau\int d^2z\,
\big\langle b\,\widetilde b\,
U^{(0,0)}(z,\bar z)\,\mathcal X(w)\,
\widetilde{\mathcal X}(\widetilde w)\,V^{(-1,-1)}(0)\big\rangle .
\end{equation}
The vertex at $z$ is integrated and the one at the origin is unintegrated.
With one PCO in each chirality, two PCOs of the same chirality cannot
collide. We move the left-moving PCO from $v$ to $w$ and hold the
right-moving PCO fixed. Since
$\mathcal X(w)-\mathcal X(v)=\{Q_B^L,\Lambda\}$ with
$\Lambda=\xi(w)-\xi(v)$, the change is
\begin{equation}
\Delta\mathcal A=\int d^2\tau\,d^2z\,
\big\langle b\,\widetilde b\,\{Q_B^L,\Lambda\}\,
\widetilde{\mathcal X}\,U^{(0,0)}V^{(-1,-1)}\big\rangle .
\label{A.1}
\end{equation}

To move the BRST contour, we use
\begin{equation}
\{Q_B^L,b\}=T,\qquad
[Q_B^L,\widetilde{\mathcal X}]=[Q_B^L,V^{(-1,-1)}]=0 .
\label{a.2}
\end{equation}
Write $U^{(0,0)}=U^{(0)}\widetilde U^{(0)}$, where
$U^{(-1)}=e^{-\phi}\mathcal O$, $U^{(0)}=G_{-1/2}\mathcal O$,
and $\mathcal O$ is a matter superconformal primary of weight $1/2$.
Only $cT_m+\gamma T_F$ in the BRST current has a singular OPE with
$U^{(0)}$, where $\gamma=\eta e^\phi$. The two OPEs are
\begin{align*}
T_m(u)U^{(0)}(z)&\sim
 \frac{U^{(0)}(z)}{(u-z)^2}+\frac{\partial U^{(0)}(z)}{u-z},\\
T_F(u)U^{(0)}(z)&\sim
 \frac{\mathcal O(z)}{(u-z)^2}+\frac{\partial\mathcal O(z)}{u-z}.
\end{align*}
The second line follows from
$G_{1/2}G_{-1/2}\mathcal O=2L_0\mathcal O=\mathcal O$ and
$G_{-1/2}^2\mathcal O=L_{-1}\mathcal O=\partial\mathcal O$.
Taking the residue gives
\begin{align}
[Q_B^L,U^{(0)}]
 &= (\partial c)U^{(0)}+c\,\partial U^{(0)}
   +(\partial\gamma)\mathcal O+\gamma\,\partial\mathcal O
  =\partial_z V^{(0)},\qquad
 V^{(0)}=cU^{(0)}+\gamma\mathcal O ,\nonumber\\
[Q_B^L,U^{(0,0)}]&=\partial_z\big[V^{(0)}\widetilde U^{(0)}\big].
\label{sdf}
\end{align}

On the torus the numbers of left-moving $b$ and $c$ ghosts must match.
Replacing $b$ by $T$ leaves the $c$ in $V^{(-1,-1)}$ unpaired, so the
stress-tensor term vanishes. The contour deformation in \eqref{A.1} then gives
\begin{equation}
\Delta\mathcal A=\int d^2\tau\,d^2z\,
\big\langle b\,\widetilde b\,\Lambda\,\widetilde{\mathcal X}\,
\partial_z\big[(cU^{(0)}+\gamma\mathcal O)\widetilde U^{(0)}\big]\,
V^{(-1,-1)}\big\rangle .
\label{eq:pco_variation_before}
\end{equation}
The $cU^{(0)}$ term has two $c$ ghosts and one $b$ ghost, so it also
vanishes. For fixed PCO positions, this leaves
\begin{align}
\Delta\mathcal A&=\int d^2\tau\,d^2z\,\partial_z F_{w,v},
\label{eq:pco_total_derivative}\\
F_{w,v}&=\big\langle b\,\widetilde b\,\Lambda\,\widetilde{\mathcal X}\,
\eta e^\phi\mathcal O(z)\,\widetilde U^{(0)}(\bar z)\,
V^{(-1,-1)}(0)\big\rangle .
\label{eq:pco_boundary_F}
\end{align}
Let $D_\tau$ be the regulated domain of $z$ at fixed $\tau$.
With $d^2z=\tfrac{i}{2}dz\wedge d\bar z$, Stokes' theorem gives
\begin{equation}
\Delta\mathcal A
 =\frac{i}{2}\int d^2\tau\oint_{\partial D_\tau}F_{w,v}\,d\bar z
 =\int_{\partial\mathcal M_{1,2}}\Theta_{w,v},
\qquad
\Theta_{w,v}=\frac{i}{2}d^2\tau\wedge F_{w,v}\,d\bar z .
\label{sed1}
\end{equation}

For an even spin structure $s=2,3,4$, we insert $\xi(0)$ to absorb
the $\xi$ zero mode. Up to a common normalization, the superghost factor is
\begin{align}
\mathcal K_s(a,z)
 &=\big\langle\xi(0)\xi(a)\eta(z)e^{\phi(z)}e^{-\phi(0)}\big\rangle_s
 =\frac{\theta_s(a-z)\theta_1(a)}{\theta_s(a)\theta_1(a-z)}
 =\frac{S_s(a-z)}{S_s(a)},\label{eq:ghost_kernel}\\
\mathcal K_s(w,v;z)
 &=\frac{S_s(w-z)}{S_s(w)}-\frac{S_s(v-z)}{S_s(v)} .
\label{eq:ghost_difference}
\end{align}
Here $S_s(z)=\theta_s(z)\theta_1'(0)/[\theta_s(0)\theta_1(z)]$
is the Szeg\H{o} kernel. At fixed loop momentum $\ell$, let
$C_s(z)=\langle\mathcal O_1(z)\mathcal O_2(0)\rangle_{s,\ell}$.
The left-moving spin sum and the full boundary correlator are
\begin{align}
f_{w,v}(z;\tau,\ell)
 &=\sum_{s=2}^4\omega_s\,\mathcal K_s(w,v;z)C_s(z),
 &\omega_s&=\frac{c_s\,\theta_s(0|\tau)^4}{2\theta_1'(0|\tau)^4},
 \label{eq:f_pco}\\
F_{w,v}
 &=\mathcal G_{bc}(\tau)\int d^{10}\ell\,
 f_{w,v}(z;\tau,\ell)\,\widetilde B(\widetilde w,\bar z;\tau,\ell).
 \label{eq:F_explicit}
\end{align}
Here $c_s$ are the GSO signs, $\mathcal G_{bc}$ is the $bc$-ghost
factor, and $\widetilde B$ is the right-moving block. Common
normalizations and bosonic zero-mode factors are included in the blocks.

For these DDF states, $C_s(z)=S_s(z)H(z)$, where $H$ is independent
of $s$ and regular at $z=0$. As $z\to0$, both ratios in
\eqref{eq:ghost_difference} approach one. Their difference is $O(z)$
and cancels the pole $S_s(z)\sim1/z$. At $z=w$ and $z=v$, the
residues of $\mathcal K_s C_s$ are $-H(w)$ and $+H(v)$, respectively.
Each is independent of $s$, so each cancels by $\sum_{s=2}^4\omega_s=0$.
The spin-summed right-moving block is also regular at these points.
Thus each collision circle gives
\begin{equation}
\lim_{\epsilon\to0}\oint_{|z-z_*|=\epsilon}F_{w,v}\,d\bar z=0,
\qquad z_*=0,w,v .
\label{eq:pco_collision_vanish}
\end{equation}

In the odd spin structure, all ten $\psi^\mu$ are periodic around both
torus cycles. The integral $\int d^{10}\psi_0$ therefore requires ten
independent fermion zero modes. For our two $(-1)$-picture vertices,
the fermions lie along at most five directions,
$\{q,\lambda,\lambda^*,\nu,\nu^*\}$. Here $q$ is the null DDF
reference vector, orthogonal to the four polarizations. The pair
$\lambda,\lambda^*$ labels the oscillator polarizations and
$\nu,\nu^*$ the NS seed polarizations. Picture changing one vertex
adds at most one $\psi$ through the matter supercurrent. The
supercurrent in the remaining PCO adds at most one more. Since
$5+1+1=7<10$, the odd contribution vanishes in either chirality.

Finally, we choose PCO positions on each patch of moduli space away
from zeros of the superghost theta functions. Vertical integration
cancels the terms at the boundaries between patches. Together with
\eqref{eq:pco_collision_vanish} and the odd-spin count, this gives
$\Delta\mathcal A=0$ for the DDF states considered here.

A collision can contribute in other backgrounds. In Sen's heterotic
Calabi--Yau example, the matter OPE for two charged massless scalars
contains a $1/\bar z$ term. Its boundary integral is proportional to
the one-loop Fayet--Iliopoulos term for an anomalous $U(1)$
\cite{Sen:2014pia}. The corresponding correction vanishes in flat
spacetime, including toroidal compactifications \cite{Sen:2016gqt}.

\end{document}